\newif\ifonecol
\onecolfalse

\documentclass{aa}

\usepackage[colorlinks=true,linkcolor=blue,citecolor=blue]{hyperref}

\usepackage{natbib}
\usepackage{txfonts}
\usepackage{graphicx}
\usepackage{mwe}

\hypersetup{colorlinks,allcolors=blue}

\def\lum        {\ensuremath{L_\text{1.4~GHz}}}

\def\whz        {\ensuremath{\text{W}\,\text{Hz}^{-1}}}
\def\mujybeam {\ensuremath{\mu\text{Jy}\,\text{beam}^{-1}}}

\def\vmax       {\ensuremath{V_\text{max}}}

\authorrunning{Vardoulaki et al.}
\titlerunning{RLF of galaxy groups in COSMOS}

\usepackage{color}
\usepackage{amstext}

\begin{document}


\title{The evolution of the radio luminosity function of group galaxies in COSMOS}

\author{
            E.~Vardoulaki\inst{1,2}\thanks{email: elenivard@gmail.com}\texorpdfstring{\href{https://orcid.org/0000-0002-4437-1773}{\protect\includegraphics{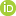}}}{0000-0002-4437-1773}, 
		  G.~Gozaliasl\inst{3,4}\texorpdfstring{\href{https://orcid.org/0000-0002-0236-919X}{\protect\includegraphics{ORCID-iD_icon-16x16.png}}}{0000-0002-0236-919X},
    A.~Finoguenov\inst{4}\texorpdfstring{\href{https://orcid.org/0000-0002-4606-5403}{\protect\includegraphics{ORCID-iD_icon-16x16.png}}}{0000-0002-4606-5403},
    M.~Novak\inst{5}\texorpdfstring{\href{https://orcid.org/0000-0002-0236-919X}{\protect\includegraphics{ORCID-iD_icon-16x16.png}}}{0000-0001-8695-825X}, 
        H. G. Khosroshahi\inst{6} \texorpdfstring{\href{https://orcid.org/0000-0003-0558-8782}{\protect\includegraphics{ORCID-iD_icon-16x16.png}}}{0000-0003-0558-8782}
        }

\institute{
        IAASARS, National Observatory of Athens,  Lofos Nymfon, 11852 Athens, Greece
        \and
Th\"{u}ringer Landessternwarte, Sternwarte 5, 07778 Tautenburg, Germany	
	\and
 Department of Computer Science, Aalto University, PO Box 15400, Espoo, FI-00076, Finland
 \and
Department of Physics, University of Helsinki, P. O. Box 64, FI-00014, 
Helsinki, Finland 
\and
 Max-Planck-Institut f\"{u}r Astronomie, K\"{o}nigstuhl 17, D-69117, Heidelberg, Germany
\and
School of Astronomy, Institute for Research in Fundamental Sciences (IPM), P.O. Box 1956836613, Tehran, Iran
}

\date{Received ; accepted 02/05/2024}

\abstract{
To understand the role of the galaxy group environment on galaxy evolution, we present a study of radio luminosity functions (RLFs) of group galaxies based on the Karl G. Jansky Very Large Array-COSMOS 3 GHz Large Project. The radio-selected sample of 7826 COSMOS galaxies with robust optical/near-infrared counterparts, excellent photometric coverage, and the COSMOS X-ray galaxy groups ($M_{200c} > 10^{13.5} M_{\odot}$) enables us to construct the RLF of group galaxies (GGs) and their contribution to the total RLF since $z \sim$ 2.3. 
Using the Markov chain Monte Carlo algorithm, we fit a redshift-dependent pure luminosity evolution model and a linear and power-law model to the luminosity functions. We compare it with past RLF studies from VLA-COSMOS on individual populations of radio-selected star-forming galaxies (SFGs) and galaxies hosting active galactic nuclei (AGN). These populations are classified based on the presence or absence of a radio excess concerning the star-formation rates derived from the infrared emission. 
We find that the density of radio galaxies in groups is low compared to the field at $ z \sim$ 2 down to $z \sim$ 1.25, followed by a sharp increase at $z \sim$ 1 by a factor of 6, and then a smooth decline towards low redshifts. This trend is caused by both decrease in the volume abundance of massive groups at high-$z$ and the changes in the halo occupation of radio AGN, which are found by other studies to reside at smaller halo mass groups. This indicates that the bulk of high-$z$ $log_{10}(M_{\rm 200c}/\rm M_{\odot}) >$ 13.5 groups must have been forming recently, and so the cooling has not been established as yet. The slope of the GG RLF is flatter compared to the field, with excess at high radio luminosities.
The evolution in the GG RLF is driven mainly by satellite galaxies in groups. 
At $z \sim$ 1, the peak in the RLF, coinciding with a known overdensity in COSMOS, is mainly driven by AGN, while at $z > 1$ SFGs dominate the RLF of group galaxies. 
A drop in occurrence of AGN in groups at $z>1$ by a factor of 6, manifests an important detail on the processes governing galaxy evolution. 
}


\maketitle

\section{Introduction}
\label{sec:intro}

The properties and evolution of galaxies are known to be strongly linked to their external environment. Massive halos are found to play a key role in galaxy evolution. At low redshifts, it has been found that clusters of galaxies are mostly dominated by early-type galaxies composed of old stellar populations, while low-density environments host typically late-type galaxies with younger and bluer stars, producing the star-formation (SF) - density - distance to cluster centers relations and affecting the morphology of galaxies \citep[e.g.,][]{oemler1974,dressler1980}. We expect that galaxies in dense regions experience various physical processes such as tidal forces, mergers, high-speed interactions, harassment, and gas stripping, which in turn contribute to dramatic morphological changes and quenching of star formation \citep[e.g.][]{larson1980,byrd1990}. However, these physical processes' precise timing and relative importance are not yet well understood.

The environmental processes which affect galaxy evolution could directly or indirectly influence the accretion onto the central black hole in galaxies, notably those with a stellar bulge \citep{magorrian1998}. Both local and large-scale processes which may affect cluster galaxies also have the potential to affect the gas distribution in the galaxies and hence may trigger or suppress active galactic nuclei (AGN) activity. 

Apart from the role of galaxy group and cluster environment on radio emission of the brightest galaxy of the group, \cite{khosroshahi2017} suggested that the radio luminosity of the brightest group galaxy (BGG) also depends on the group dynamics, in a way that BGGs in groups with a relaxed/virialised morphology are less radio luminous than the BGG with the same stellar mass but in an evolving group. This was supported numerically by a semi-analytic approach \citep{raouf2018}, where they predicted the radio power for the first time. However, the numerical models cannot be constrained without an observational constraint reaching high redshift.  

Many radio studies \citep{best2002,barr2003,miller2003,reddy2004} showed an increase in radio-loud AGN activity in galaxy clusters, at a range of redshifts, and in both relaxed and merging systems. The radio emission ($<30$ GHz) in galaxies is dominated by synchrotron radiation from accelerating relativistic electrons, with a fraction of free-free emission \citep[e.g.,][]{sadler1989,condon1992,clemens08,tabatabaei2017}. The feedback from supernovae explosions in star-forming galaxies (SFGs) and that from the growth of the central supermassive black hole (SMBH) in AGN are two main sources of acceleration of cosmic electrons.

To use radio emission as a proxy for measuring star formation rates (SFRs) or AGN feedback, it is important to estimate which process dominates the radio emission: star formation processes or SMBH accretion. We follow the method demonstrated in \citep{delvecchio17} who measured the radio excess compared to the total star-formation-based infrared (IR) emission. Objects which exhibit radio excess above what is expected from star formation alone, as calculated from their infrared emission, are deemed AGN, and the rest are SFGs. These populations contribute different percentages to the energy budget. In the radio, this is quantified by calculating the radio luminosity function (RLF). \cite{novak18} studied the 3 GHz VLA-COSMOS RLF and calculated the relative contributions to the RLF from the AGN and SFG populations down to submicrojansky levels. AGN and SFGs contribute differently to the RLF, where AGN are known to dominate the bright part of the RLF, and SFGs dominate the faint. In particular, 90\% of the population at the faint end ($< 0.1$mJy) is linked to SFGs. In clusters of galaxies, \cite{yuan16} who studied the RLF of brightest cluster galaxies (BCGs) up to $z$ = 0.45 found no evolution, and a dominant population of AGN, as most of their BCGs are associated with AGN. \cite{branchesi06} compared clusters at 0.6 $< z < 0.8$ to the local Abell clusters and found very different RLFs. These studies target populations dominated by AGN and thus probe the high end of the radio luminosity function. The question arises, how much do smaller mass environments, those of groups of galaxies, and their members contribute to the observed radio source population.

In this paper we investigate the population of galaxies inside X-ray galaxy groups in the COSMOS field \citep{gozaliasl19} to quantify their contribution to the RLF at 3 GHz VLA-COSMOS \citep{novak18,smolcic17a}. Section~\ref{sec:data} describes the X-ray and radio data used throughout this work. Section~\ref{sec:lumfun} focuses on methods for deriving the RLF and its evolution through cosmic time. In Section~\ref{sec:evolution}, we present and discuss the results on the RLF of group galaxies and calculate their contribution to the total RLF at 3 GHz. We further separate the galaxies to BGGs and satellites (SGs). We also use the radio excess parameter and the presence of jets/lobes to disentangle AGN and SFGs in the radio and provide the relative contributions of these populations to the group galaxies (GG) RLF and the total RLF. This is presented in Section~\ref{sec:agn}.  Finally, in Section~\ref{sec:summary}, we provide a summary. The tables with the analysis results can be found in the Appendix.

We assume flat concordance Lambda Cold Dark Matter ($\Lambda$CDM) cosmology defined with a Hubble constant of  $H_0=70$\,km\,s$^{-1}$\,Mpc$^{-1}$, dark energy density of $\Omega_{\Lambda}=0.7$, and matter density of $\Omega_\text{m}=0.3$. For the radio spectral energy distribution, we assume a simple power law described as $S_{\nu}\propto\nu^{-\alpha}$, where $S_\nu$ is the flux density at frequency $\nu$ and $\alpha$ is the spectral index. If not explicitly stated otherwise, $\alpha=0.7$ is assumed.

\section{The Data}
\label{sec:data}

The Cosmic Evolution Survey (COSMOS) is a deep multi-band survey covering a 2 deg$^2$ area, thus offering a comprehensive data set to study the evolution of galaxies and galaxy systems. The full definition and survey goals can be found in \cite{scoville07}. The sample selection for this study is described below.

\subsection{Radio selected galaxies}

 We used radio-selected samples of galaxies cross-matched with multi-wavelength optical/near-infrared (NIR) and value-added catalogues in the COSMOS field. The radio data have been selected from the VLA-COSMOS 3~GHz Large Project \citep{smolcic17a}, with a median sensitivity of  2.3~\mujybeam\ and resolution of $0.75$  arcsec. The cross-correlation of the radio and multiwavelength sources can be found in \cite{smolcic17b}. Only sources within the COSMOS2015 catalogue \citep{laigle16} or with $i$-band counterparts have been given the availability of reliable redshift measurements. The COSMOS2015 catalogue contains the high-quality multiwavelength photometry of $\sim$800\,000 sources across more than 30 bands from near-ultraviolet (NUV) to near-infrared (NIR) through several surveys and legacy programs \citep[see][ for detailed description]{laigle16}.

\subsection{X-ray galaxy groups catalogue}
\label{sec:xraydata}
 
 \cite{finoguenov07} and \cite{george11} presented primary catalogs of the X-ray galaxy groups in COSMOS. These catalogs combined the available \textit{Chandra} and \textit{XMM-Newton} data (with improvements in the photometric datasets) used to identify galaxy groups, with secure identification reaching out to $z\sim 1.0$. 
 On completion of the visionary \textit{Chandra} program \citep{elvis09, civano16}, high-resolution imaging across the full COSMOS field became available. Furthermore, more reliable photometric data provided a robust identification of galaxy groups at a higher redshift, thus resulting in a revised catalogue of extended X-ray sources in COSMOS \citep{gozaliasl19}, which was obtained by combining both the \textit{Chandra} and \textit{XMM-Newton} data for the COSMOS field. 
 
 The COSMOS galaxy group catalogue that we use in this study relies on a combination of an updated version of the initial group catalogs with 183 groups and a new catalogue of 73 groups described in \cite{gozaliasl19} and Gozaliasl et al. (in preparation), which combines data of all X-ray observations from \textit{Chandra} and \textit{XMM-Newton} in the 0.5-2 keV band, with robust group identification up to $z\sim 2.0$. It reaches an X-ray limit of $3\times10^{-16}~erg~cm^{-2}~s^{-1}$ in the range 0.5-2 keV and contains groups with $M_{\rm 200c} = 8\times10^{12}-3\times10^{14}M_{\odot}$.

 Group halo mass is the total mass (commonly called $M_{200c}$) which was determined using the scaling relation $L_{X}-M_{200c}$ with weak lensing mass calibration as presented by \citet{leauthaud10}. The radius of the group $R_{200}$ is defined as the radius enclosing $M_{200c}$ with a mean overdensity of $\Delta\sim 200$ times the critical background density. \cite{gozaliasl19} discussed the mass completeness of the group sample given the surface brightness limitation of the X-ray dataset. Over the redshift range $0.5<z<1.2$, the evolution of the group mass limit is weak and lies within the observational uncertainties, being around $\log (M_{\rm group}/{\rm M}_{\odot}) \sim 13.38$ at $z\sim 0.5$ and  $\log (M_{\rm group}/{\rm M}_{\odot}) \sim 13.5$ at $z > 0.5$.

 The redshift of the group is the redshift of the peak of the galaxy distribution within the group radius while slicing the lightcone with a redshift step of 0.05. In most cases, this redshift determination is strengthened by the presence of spectroscopic galaxies redshifts. The brightest group galaxy (BGG) is detected from the COSMOS2015 photometry as being the most massive galaxy within $R_{200}$, with a redshift that matches that of the hosting group \citep{gozaliasl19}. More than $\sim80\%$ of the BGGs have robust spectroscopic redshifts. The center of groups from the X-ray emission is determined with an accuracy of $\sim 5\arcsec$, using the smaller scale emission detected by \textit{Chandra} data. The BGGs do not always locate at the peak of the X-ray center emission. As described in \cite{gozaliasl19}, the off-central BGG probably resides in groups more likely to have experienced a recent halo merger. The rest of the group galaxies (GGs) are called satellites (SGs).
 
\begin{figure}
\centering
\includegraphics[width=\linewidth]{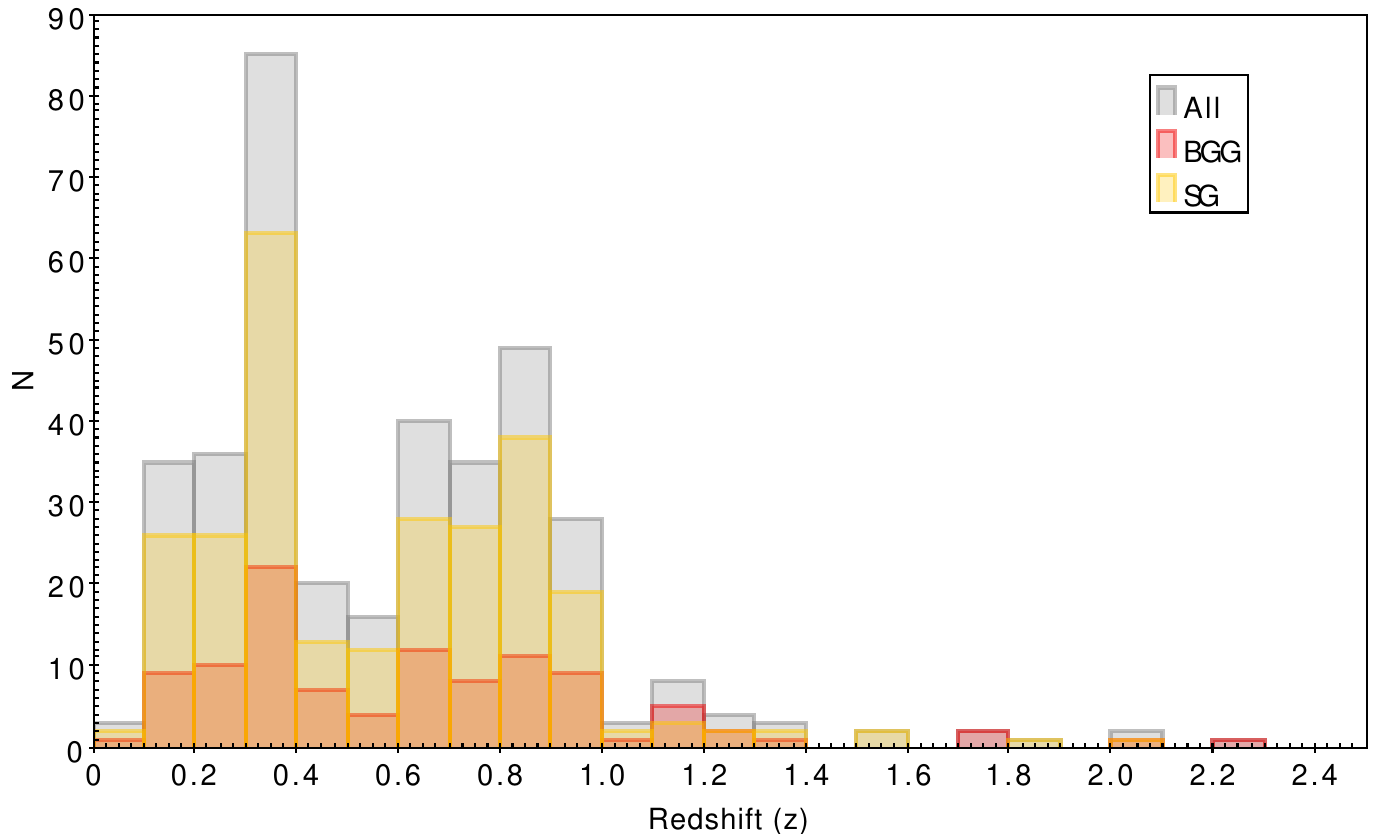}
\includegraphics[width=\linewidth]{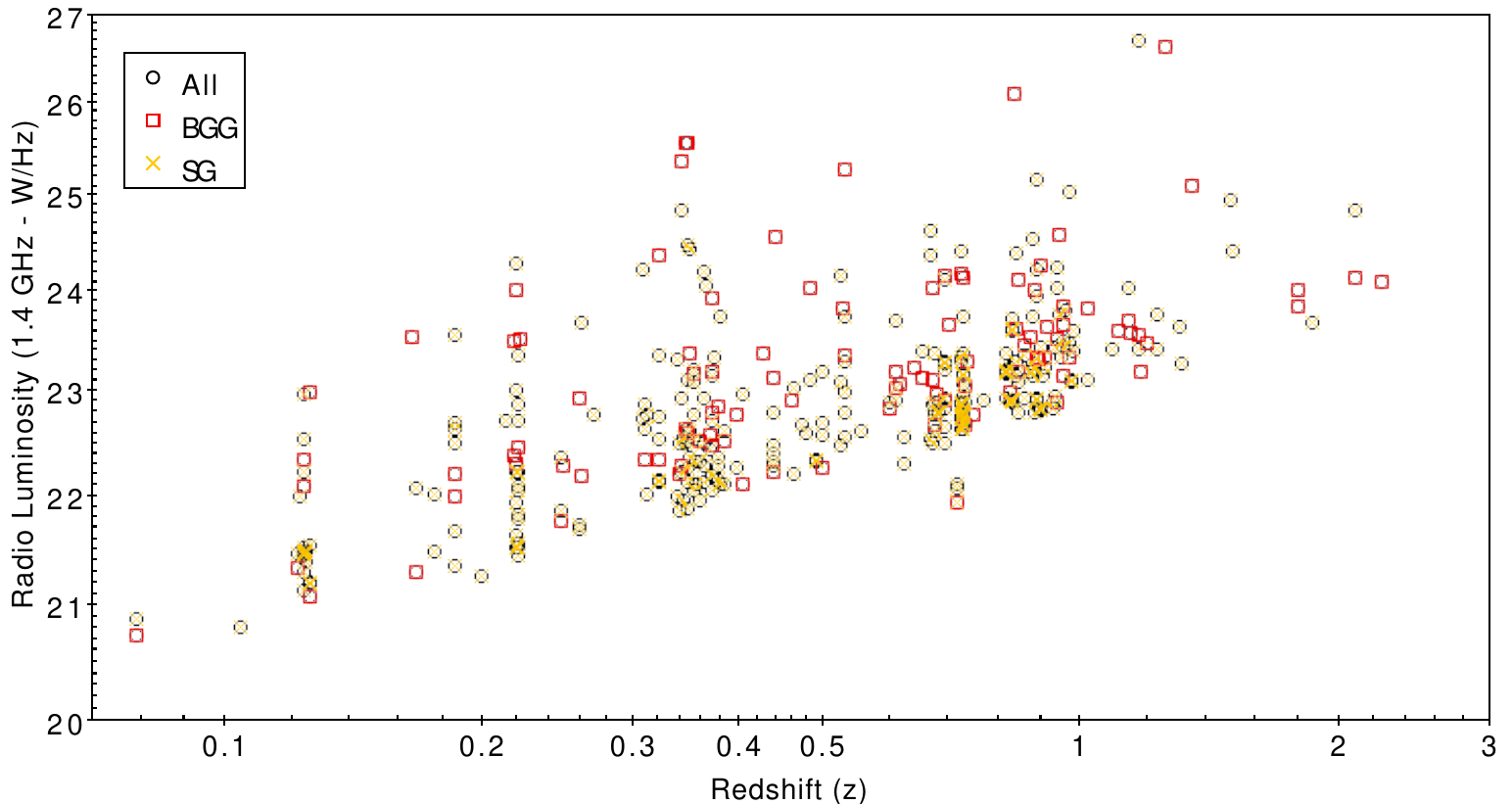}
\includegraphics[width=\linewidth]{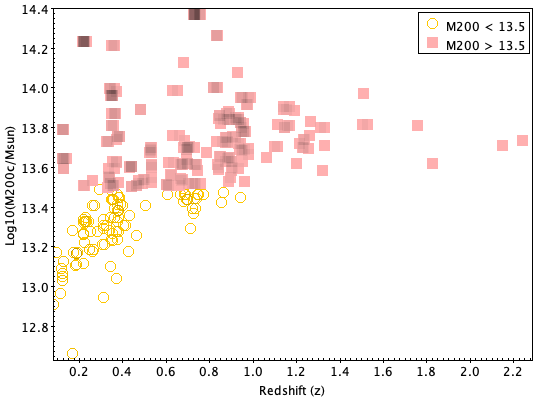}
\caption{Top: Number of sources per redshift. The bin size is 0.1. Middle: Radio luminosity at 1.4 GHz versus redshift. The redshift plotted is the one of the galaxy groups. The radio luminosity is calculated from the 1.4 GHz flux density for the redshift of the object. Black represents all group galaxies, red is for BGGs, and yellow is for SGs (see Sec.~\ref{sec:data} for clarification on the classification). Bottom: Halo mass versus redshift. Pink filled squares denote log10($M_{200}/M_{\odot}) > 13.5$ and yellow open circles log10($M_{200}/M_{\odot}) < 13.5$, which we refer to as log10($M_{200}/M_{\odot}) \approx 13.3$ in the rest of the paper. The divide shows our adopted halo mass cut to account for sample completeness (Sec.~\ref{sec:xraydata}).}
\label{fig:z_gg}
\end{figure}

 A quality flag has been assigned to groups depending on the robustness of the extraction and the potential availability of spectroscopic redshift \citep{gozaliasl19}. In our study, we keep only the groups with flags 1, 2, and 3. We considered only groups with BGG galaxies more massive than $\log M_{*}/{\rm M}_{\odot}=10$. We refer the reader to \cite{gozaliasl19} for further information on identifying groups.
 Within the virial radius of these groups, the above selection criteria resulted in a total of 306 objects distributed in the galaxy groups. In Fig.~\ref{fig:z_gg} we present the data for the group galaxies used in our analysis. The spectroscopic redshifts are available for 35\% of our sources, and the median accuracy of the photometric redshifts is $\Delta z/(1+z_\text{spec})= 0.007$ \citep{laigle16}.

\subsection{Sample of group galaxies used in this analysis}

To analyse the radio luminosity function of group galaxies in COSMOS, we cross-match the galaxy group catalogue and the 3 GHz VLA-COSMOS data \citep{smolcic17a} within a radius of $0.8{\arcsec}$. We furthermore use the 3 GHz VLA-COSMOS data presented in \cite{novak18}, who constructed RLFs up to $z \sim$ 5.5, to compare to the total RLF in COSMOS up to $z \sim$ 2.3. Additionally, \cite{novak18} separated objects in SFGs and AGN, following the radio excess prescription of \cite{delvecchio17}. This method is based on the excess radio emission expected from star formation alone. \cite{delvecchio17} fitted the infrared spectral energy distribution of radio sources at 3 GHz VLA-COSMOS and calculated the contribution of the 3 GHz VLA-COSMOS radio sources to the radio luminosity by applying a conservative cut. Galaxies that exhibit an excess in radio emission above $3\sigma$ from what is expected from SF alone were deemed AGN, with the rest being SFGs. This method was used to separate the \cite{novak18} sample, which we use here for comparison, in SFGs and AGN. Finally, \cite{novak18} described possible biases and uncertainties associated with the data sample selection, and thus we refer the reader to this reference.

\section{Methods and analysis}
\label{sec:lumfun}

We describe the process of calculating the RLF for galaxies in groups in COSMOS \citep{gozaliasl19} using the VLA-COSMOS 3~GHz data.  We applied a cut in group mass $M_{200c} > 10^{13.5} M_{\odot}$, to account for a difference in the limiting mass of the group catalogue with redshift (Sec.~\ref{sec:xraydata}). This cut is demonstrated at the bottom frame of Fig.~\ref{fig:z_gg}. We further separate group galaxies in BGGs and SGs. We compare the RLF of the population of SFGs and AGN at 3 GHz VLA-COSMOS in the same redshift bins to the total RLF calculated from the 3 GHz data \citep{novak17, smolcic17a}. We fit linear and power-law models to the RLFs of GGs and compare them to the total RLF to obtain the contribution of GGs to the total RLF at 3 GHz VLA-COSMOS, something that has not been shown before in COSMOS.

\subsection{Measuring the radio luminosity function}
\label{sec:lumfun_total}

To obtain the total RLFs, for GGs, SFGs, and AGN, we followed the method adopted by \cite{novak17} (see their Sec. 3.1). They computed the maximum observable volume \vmax\, for each source \citep{schmidt68} and simultaneously applied completeness corrections that take into account the non-uniform $rms$ noise and the resolution bias \citep[see Sec.~3.1 in][]{novak17}.   
Then the RLF is
\begin{equation}
\Phi(L,z) = \frac{1}{\Delta {\rm log} L} \sum_{i=1}^{N} \frac{1}{V_{max,i}}
\label{eq:phi}
\end{equation}

\noindent
where $L$ is the rest-frame luminosity at 1.4 GHz, derived using the radio spectral index of a source calculated between 1.4 GHz \citep{schinnerer10} and 3~GHz \citep{smolcic17a}, and $\Delta log L$ is the width of the luminosity bin. The radio spectral index should remain unchanged between frequencies and is only available for a quarter of the 3 GHz VLA-COSMOS sample. For the rest of the sources detected only at 3~GHz, we assumed $\alpha=0.7$. The latter corresponds to the average spectral index of the entire 3~GHz population \citep[see Sec.~4 in][]{smolcic17a}. $V_{\rm max}$ is the maximum observable volume given by

\begin{equation}
V_{max,i} =  \sum_{z=z_{min}}^{z_{max}} [V(z + \Delta z) - V(z)] C(z), 
\label{eq:vmax}
\end{equation}

\noindent
where the sum starts at $z_{min}$ and adds co-moving spherical shells of volume $\Delta V = V(z + \Delta z) - V(z)$ in small redshift steps $\Delta z$ = 0.01 until $z_{max}$. $C(z)$ is the redshift-dependent geometrical and statistical correction factor. This takes into account sample incompleteness. For a thorough description of the biases, see Section~6.4 in \cite{novak17}. The correction factor is given by
\begin{equation} 
C(z) = \frac{A_{obs}}{41 253 deg^{2}} \times C_{radio}(S_{3 GHz}(z)) \times C_{opt}(z),
\label{eq:cz}
\end{equation}

\noindent
where $A_{obs}$ = 1.77 deg$^{2}$ is the effective unflagged
area observed in the optical to NIR wavelengths, $C_{radio}$ is the
completeness of the radio catalogue as a function of the flux density $S_{3 GHz}$, and $C_{opt}$ is the completeness owing to radio sources
without assigned optical-NIR counterpart. Completeness corrections are shown in \cite{smolcic17a} in their Fig. 16 and Table 2, and in \cite{novak17} in their Fig. 2.

\begin{figure*}
 \centering
 \resizebox{\hsize}{!}
  {
 \includegraphics[width=\linewidth]{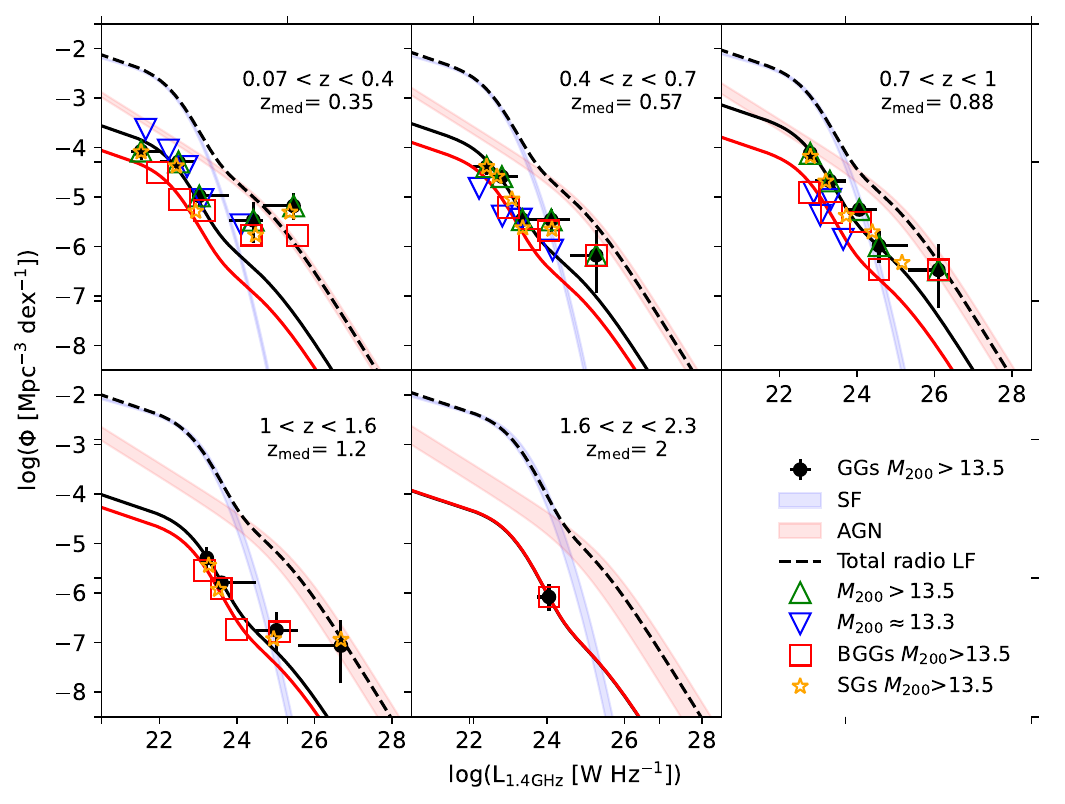}
 }
 \caption{Total radio luminosity functions of galaxies in groups. Black points indicate the RLFs for group galaxies derived using the \vmax\ method (see Section~\ref{sec:lumfun_total}). Red squares and yellow stars mark the brightest group galaxies and satellites, respectively. The blue and red shaded areas show the $\pm 3\sigma$ ranges of the best-fit evolution for the individual SFG and AGN populations, respectively (outlined in Section~\ref{sec:lumfun_pop}). The black dashed line is the fit to the total RLF at 3 GHz VLA-COSMOS \citep{novak17, smolcic17c}. 
 For the $z <1$ sub-samples ($z <$ 0.4), the halos have been split into massive ($M_{\rm 200c} > 10^{13.5} M_{\odot}$) and low-mass halos ($M_{\rm 200c} \approx 10^{13.3} M_{\odot}$); the latter are shown for completion but not used in the analysis. For the rest of the redshift bins, all samples have $M_{\rm 200c} > 10^{13.5} M_{\odot}$. A halo mass cut, $M_{\rm 200c} > 10^{13.5} M_{\odot}$, was applied to the GGs, BGGs, and SGs. The black solid line is the scaled fit to the group galaxies RLF, and the red solid line is the scaled fit for BGGs, as explained in Section~\ref{sec:lumfun_scalefit}. The scaled fit can be found in Fig.~\ref{fig:lfgrid_scale}, and we do not show it here for SGs for clarity. The black line for the scaled fit for GGs in the last redshift bin is hidden by the red line for BGGs, because there are no SGs in that bin.} 
 \label{fig:lfgrid_scale}
 \end{figure*}

\begin{figure*}
\centering
\resizebox{\hsize}{!}
 {
\includegraphics[width=\linewidth]{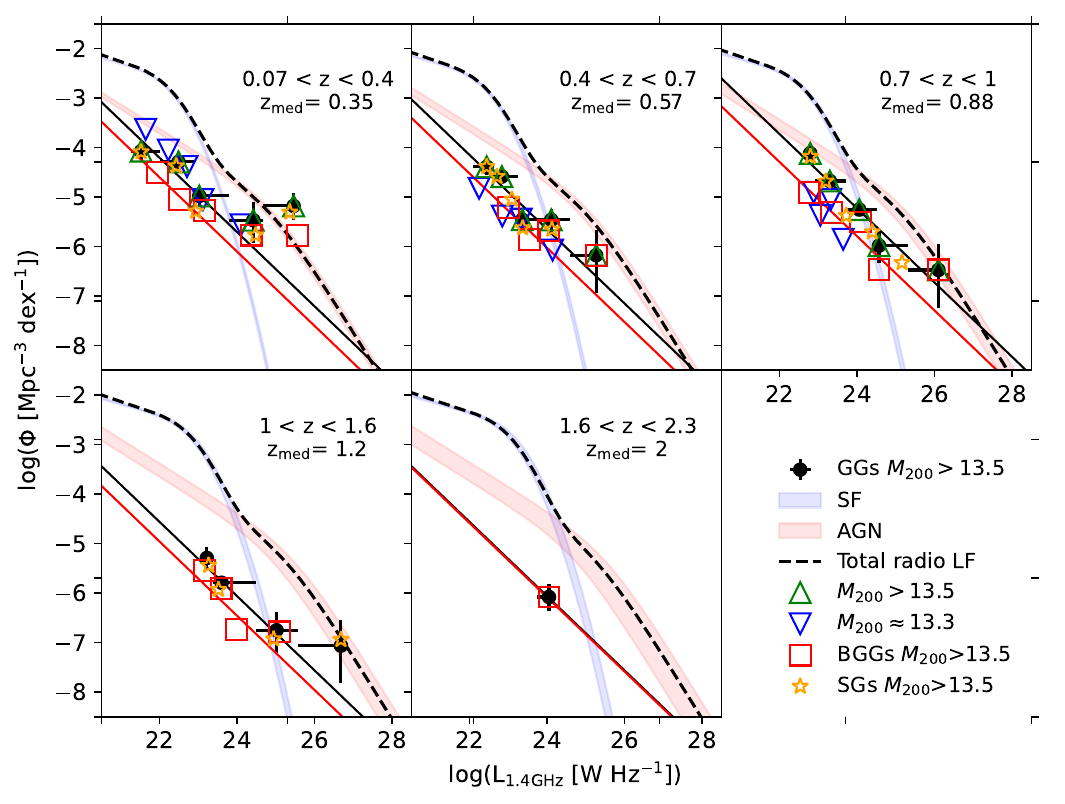}}
\caption{Same as Fig.~\ref{fig:lfgrid_scale}: Total radio luminosity functions of galaxies in groups. Black points indicate the RLFs for group galaxies derived using the \vmax\ method (see Section~\ref{sec:lumfun_total}). Red squares and yellow stars mark the brightest group galaxies and satellites, respectively. The blue and red shaded areas show the $\pm 3\sigma$ ranges of the best-fit evolution for the individual SFG and AGN populations, respectively (outlined in Section~\ref{sec:lumfun_pop}). The black dashed line is the fit to the total RLF at 3 GHz VLA-COSMOS \citep{novak17, smolcic17c}. For the $z <1$ sub-samples ($z <$ 0.4), the halos have been split into massive ($M_{\rm 200c} > 10^{13.5} M_{\odot}$) and low-mass halos ($M_{\rm 200c} \approx 10^{13.3} M_{\odot}$); the latter are shown for completion but not used in the analysis. For the rest of the redshift bins, all samples have $M_{\rm 200c} > 10^{13.5} M_{\odot}$. A halo mass cut, $M_{\rm 200c} > 10^{13.5} M_{\odot}$, was applied to the GGs, BGGs, and SGs. The black solid line is the fit to the group galaxies RLF, and the red solid line is the best fit for BGGs, as explained in Section~\ref{sec:lumfun_linfit}. We do not show the best fit for SGs for clarity. The black line for the linear regression fit for GGs in the last redshift bin is hidden by the red line for BGGs, because there are no SGs in that bin. 
}
\label{fig:lfgrid}
\end{figure*}

The redshift bins are large enough not to be affected severely by photometric redshift uncertainty and follow the selection of \cite{novak17} to allow comparisons. Luminosity bins in each redshift bin span the data's observed luminosity range. To eliminate possible issues due to poorer sampling, the lowest luminosity ranges from  the faintest observed source to the 5$\sigma$ detection threshold at the upper redshift limit (corresponding to $5\times 2.3~\mujybeam$ at 3~GHz). The reported luminosity for each RLF is the median luminosity of the sources within the bin. The RLFs for all group galaxies are shown in Figs~\ref{fig:lfgrid_scale}~and~\ref{fig:lfgrid} (black points) and are also listed in Table~\ref{tab:lumfun_vmax}. The RLFs for the BGGs and SGs are also shown in Figs~\ref{fig:lfgrid_scale}~and~\ref{fig:lfgrid} (red squares/yellow stars) and are listed in Tables~\ref{tab:lumfun_bgg}~and~\ref{tab:lumfun_sg}, respectively. The $z$ bins in Figs~\ref{fig:lfgrid_scale}~and~\ref{fig:lfgrid} are split in two halo mass $M_{\rm 200c}$ bins, above and below $10^{13.5}M_{\odot}$, and for our further analysis, we use the values above. We note that at $1.6 < z < 2.3$ we do not have SGs above our halo-mass cut ($M_{\rm 200c} > 10^{13.5} M_{\odot}$). This is related to limits in the radio power that are probed at those redshifts, leading to low statistics and scarcity of less massive groups. As discussed in \cite{novak18}, there is only a 5-10\% loss of completeness on the optical/NIR counterparts above $z\sim2$.

\subsection{The total RLF at 3 GHz VLA-COSMOS}
\label{sec:lumfun_pop}

We use the total RLF derived from the SFG and AGN populations at 3 GHz VLA-COSMOS \citep{novak17, novak18, smolcic17c} to compare to the RLF values derived for the group galaxies in COSMOS. The RLF of the SFG and AGN populations are calculated similarly for the same redshift bins as described above and for the same area coverage as the galaxy groups in COSMOS. To fit the RLF, two models are used in literature \citep[e.g.,][]{condon84, sadler02, gruppioni13}, the pure luminosity evolution (PLE) and the pure density evolution (PDE). The RLF is fitted, assuming its shape remains unchanged at all observed cosmic times. Only the position of the turnover and normalisation can change with redshift. This corresponds to the translation of the local LF in the $log{L} - log{\Phi}$ plane \citep{condon84} and can be divided into pure luminosity evolution (horizontal shift) and pure density evolution (vertical shift). 

To describe an RLF across cosmic time, the local RLF is evolved in luminosity or density, or both \citep[e.g.][]{condon84}. This is parametrised \citep{novak18} using two free parameters for density evolution ($\alpha_{D}$, $\beta_{D}$), and two for luminosity evolution ($\alpha_{L}$, $\beta_{L}$) to obtain
\begin{equation}
\Phi(L,z, \alpha_L, \beta_L, \alpha_D, \beta_D) =  (1+z)^{\alpha_D+z\cdot\beta_D}\times\Phi_{0} \left( \frac{L}{(1+z)^{\alpha_L+z\cdot\beta_L}}
\right),
\label{eq:lfevol_model}
\end{equation}

\noindent
where $\Phi_{0}$ is the local RLF. Since the shape and evolution of the RLF depend on the galaxy population type, \cite{novak17} used a power-law plus log-normal shape of the local RLF for SFGs. They used the combined data from \cite{condon02}, \cite{best05} and \cite{mauch07} to obtain the best fit for the local value
\begin{equation}
\Phi_0^{\text{SF}}(L)=\Phi_\star\left(\frac{L}{L_\star}\right)^{1-\alpha} \exp\left[-\frac{1}{2\sigma^2}\log^2\left(1+\frac{L}{L_\star}\right)\right],
\label{eq:lflocal_sf}
\end{equation}

\noindent
where $\Phi_\star=3.55\times10^{-3}~\text{Mpc}^{-3}\text{dex}^{-1}$, $L_\star=1.85\times10^{21}~\whz$, $\alpha=1.22$, and $\sigma=0.63$.

It was noted by \cite{novak17} that the PDE of SF galaxies would push the densities to very high numbers, thus making them inconsistent with the observed cosmic star formation rate densities. This is a consequence of the fact that our data can constrain only the bright log-normal part of the SF RLF. For AGN, it was shown by \cite{smolcic17c} that the PDE and PLE models are similar, mostly because the shape of the RLF does not deviate strongly from a simple power law at the observed luminosities. Considering the above reasoning, while also trying to keep the parameter space degeneracy to a minimum, we decided to use only the PLE for our analysis. Thus we adopt the approach of \cite{novak18}, who fitted the total RLF for SFG and AGN populations by constructing a four--parameter redshift-dependent pure luminosity evolution model with two parameters for the SFG and AGN populations of the form:

\begin{equation}
\begin{split}
\Phi(&L,z,\alpha_L^{\text{SF}}, \beta_L^{\text{SF}}, \alpha_L^{\text{AGN}}, \beta_L^{\text{AGN}})  =\\ &=\quad\Phi_{0}^{\text{SF}} \left( \frac{L}{(1+z)^{\alpha_L^{\text{SF}}+z\cdot\beta_L^{\text{SF}}}}\right)+
\Phi_{0}^{\text{AGN}} \left( \frac{L}{(1+z)^{\alpha_L^{\text{AGN}}+z\cdot\beta_L^{\text{AGN}}}}\right),
\label{eq:lf_total}
\end{split}
\end{equation}

\noindent
where $\Phi_{0}^{\text{SF}}$ is the local RLF for SFGs as in Eq.~\ref{eq:lflocal_sf}, and for the non-local Universe is a function of the quantity in the parenthesis, $L / (1+z)^{\alpha_L^{\text{SF}}+z\cdot\beta_L^{\text{SF}}}$. $\Phi_{0}^{\text{AGN}}$ is the local RLF for AGN of the form
\begin{equation}
\Phi_0^{\text{AGN}}(L)=\frac{\Phi_\star}{(L_\star / L)^{\alpha} + (L_\star / L)^{\beta}},
\label{eq:lflocal_agn}
\end{equation}

\noindent
where $\Phi_\star=\frac{1}{0.4}10^{-5.5}~\text{Mpc}^{-3}\text{dex}^{-1}$, $L_\star=10^{24.59}~\whz$, $\alpha=-1.27$, and $\beta=-0.49$ \citep{smolcic17c, mauch07}, and for the non-local Universe is a function of the quantity $L / (1+z)^{\alpha_L^{\text{AGN}}+z\cdot\beta_L^{\text{AGN}}}$. 

\cite{novak18} used the Markov chain Monte Carlo (MCMC) algorithm, available in the Python package \textsc{emcee} \citep{foreman13}, to perform a multi-variate fit to the data. The redshift dependence of the total evolution parameter $\alpha+z\cdot\beta$ (see Eq.~\ref{eq:lf_total}) is necessary to describe the observations at all redshifts. The best fit values,  based on the results of \cite{novak18}, for SFGs are $\alpha_L^{\text{SF}}=3.16$ and  $\beta_L^{\text{SF}}=-0.32$, and for AGN are $\alpha_L^{\text{AGN}}=2.88$ and  $\beta_L^{\text{AGN}}=-0.84$. The $\alpha_{L}$ and $\beta_{L}$ values for both SFGs and AGN are valid for $z <$ 5.5 and within the redshift range of our sample of group galaxies. We use these values to plot the fit to the RLF for SFGs and AGN in Fig.~\ref{fig:lfgrid}, shown in blue and red lines, respectively. The total RLF (including all SFG and AGN), is shown as a dashed black line in Fig.~\ref{fig:lfgrid}.

\subsection{Fitting the RLF of group galaxies and comparing to the total 3 GHz RLF}
\label{sec:lumfun_fit}

Fitting the GG RLF is not a simple task. Performing an MCMC fit to the GG data with four free parameters ($\alpha_{\rm SF}$, $\beta_{\rm SF}$, $\alpha_{\rm AGN}$ and $\beta_{\rm AGN}$), in a similar way it was done in \cite{novak18} is proven problematic and the MCMC does not converge. Below we investigate whether the functional form of the RLF, as presented above can fit the GG RLF.

\subsubsection{Scaled fit}
\label{sec:lumfun_scalefit}

\begin{figure}
\centering
\resizebox{\hsize}{!}
 {
\includegraphics[width=\linewidth]{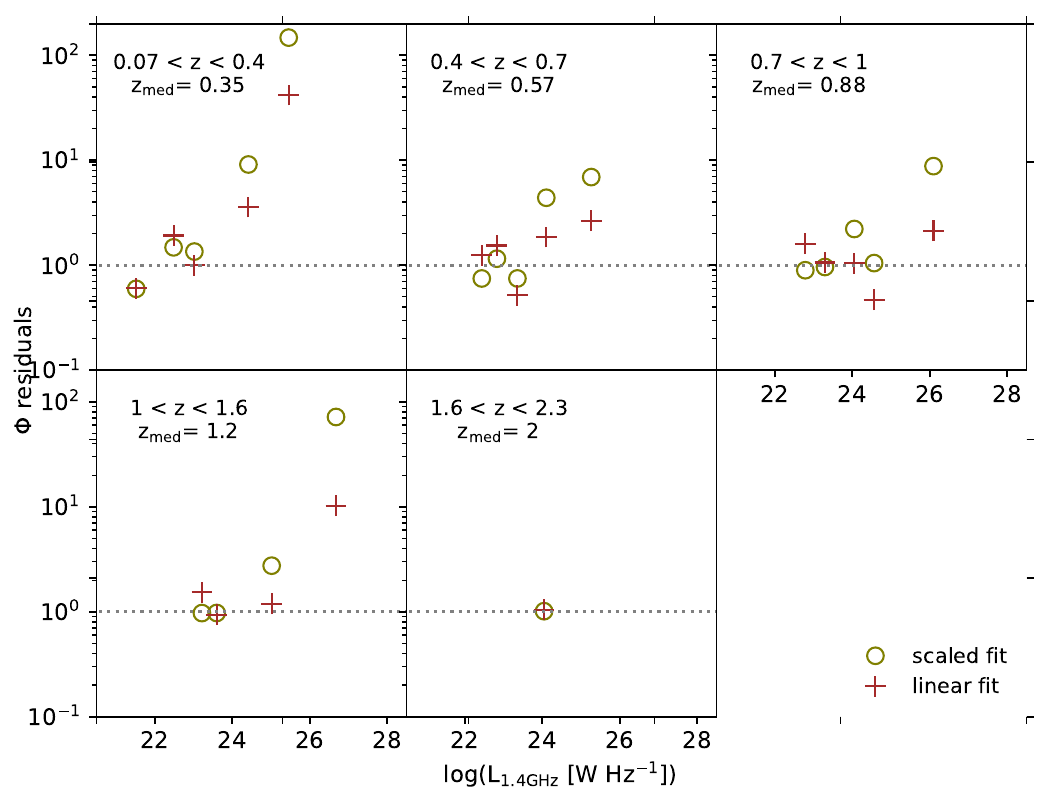}}
\caption{Ratio of $\Phi$ GG RLF data points in the adopted luminosity range and in each redshift bin, and the corresponding model value. Olive open circles denote the scaled fit method (Sec.~\ref{sec:lumfun_scalefit}) and brown crosses denote the linear regression fit method (Sec.~\ref{sec:lumfun_linfit}). The dotted grey line shows a perfect agreement between data and model.
}
\label{fig:residuals}
\end{figure}

We scale the total 3 GHz RLF to fit the GG RLF using MCMC. 
The resulting GG RLF is shown in Fig.~\ref{fig:lfgrid_scale}. There is a large deviation in the RLF values of GGs compared to the scaled fit, especially above luminosities $10^{26} \rm W~Hz^{-1}$. We perform a $\chi^{2}$ test for the goodness of the fit between the GG values and the predicted scaled fit. The results are shown in Table~\ref{tab:chi2_agn_sfg} in the Appendix. The $\chi^{2}$ is large for most redshift bins, suggesting the model is not a good fit to the data. 
 Additionally, we plot the ratio between data and model (see Fig.~\ref{fig:residuals}). It is evident that for the redshift bins with $z_{\rm med}$ = 0.35 and 1.2 the model under-fits the data in the high luminosity bins by up to $\sim$2 orders of magnitude, suggesting that the functional form of the RLF does not work well in the case of group galaxies. Hence, we explore an alternative method for fitting the GG RLF.

\subsubsection{Power-law fit}
\label{sec:lumfun_linfit}

We fit a power-law (linear regression fit in log-log) to the radio luminosity function of the group galaxies, and separately of the BGGs and SGs, from redshifts 0.07 to 2.3 of the form
\begin{equation}
y = \Phi_\star / (L_\star^\gamma) * L^\gamma, 
\label{eq:plaw}
\end{equation}

\noindent
where $L_{\star}$ is arbitrarily chosen to be $10^{24} \rm ~W ~Hz^{-1}$, and $\Phi_{\star}$ is the density at $L_{\star}$. Table~\ref{tab:lf_fit_par} shows the results for each fit. The best-fit model for the GGs is shown in Fig.~\ref{fig:lfgrid}, where we also compare the radio luminosity function of group galaxies to the total radio luminosity function of radio galaxies in the 3GHz VLA-COSMOS survey and the radio luminosity function of AGN and the star-forming population. We also present the linear regression fit for BGGs, but exclude the one for SGs for clarity. The radio emission due to the star formation over-weighs that from the AGN in lower redshifts, i.e., at $z<1$, except at high radio luminosity bins where the AGN contribution dominates. This behaviour has been described in \cite{novak17, smolcic17a} and agrees with other surveys. Adopting a fit similar to the total RLF, similar to that used in \cite{novak18}, provides a very poor fit to the GG data, particularly in the high luminosity bins. Similarly, allowing for both luminosity and $\Phi_{*}$ to be free parameters resulted in the slope fit being dominated by two points with the smallest error. These tell us the GGs do not necessarily follow the shape of the total 3GHz RLF, particularly at high luminosities, where we see increased radio activity of GG member galaxies.

\subsubsection{Methods comparison}
\label{sec:methodscomparison}

To investigate which method fits best the GG RLF, we compare the predicted to the real values. As mentioned earlier, we performed a $\chi^{2}$ test for the goodness of the fit between the GG values and the predicted (Table~\ref{tab:chi2_agn_sfg}). The test yields similar results for both methods, where the linear regression fit gives slightly better results, but neither of the models is good fit to the data.

Furthermore, we visually inspect the RLF grid plots in Figs~\ref{fig:lfgrid_scale}~\&~\ref{fig:lfgrid}. Both methods provide fairly good fits to the data. Should we use these fitted lines to predict a value, it would not necessarily agree with the observed values in some areas of the parameter space. In particular, as seen in Fig.~\ref{fig:lfgrid_scale} for the scaled RLF, the scaled RLF does not fit well the GGs above radio luminosity $10^{25}$ W/Hz in all redshift bins, i.e. the part that can be dominated by AGN contribution (based on the 3GHz RLF). In Fig.~\ref{fig:lfgrid}, for the power-law RLF, we also see the fit is not good for GGs above radio luminosity $10^{25}$ W/Hz but mainly in the redshift bins $z \sim$ 0.35 and $z \sim 1.2$. From the ratio between data and model in Fig.~\ref{fig:residuals} we see that the linear regression fit deviates less, on average, than the scaled fit from the observational data at all luminosities and redshifts.

To sum up, the $\chi^{2}$ test suggests neither of the fitting methods fits the data well. By visual inspection and from the ratio between data and model (Fig.~\ref{fig:residuals}) we see that we under-fit the GG RLF when using either of the two fitting methods for radio luminosities above $10^{25} \rm W~Hz^{-1}$ in the redshift bins $z_{\rm med}$ = 0.35 and 1.2. Below this value, there is a good rough agreement, with marginally better fits for the power-law RLF. The  
ratio between data and model in Fig.~\ref{fig:residuals} aids us in selecting a method, i.e. the power-law, linear regression fit. Thus, for the remainder of this analysis we use the power-law fit method, but also present results of the scaled method for completeness (see Table~\ref{tab:lf_fit_par_scaled}).

\subsubsection{GG contribution to the total RLF}

We calculate fractions obtained by the different methods we examined, the fraction of GG RLF to the total RLF if we apply the power-law linear regression fit (Sec.~\ref{sec:lumfun_linfit}) and the fraction of GG RLF to the total RLF if we apply the scaled RFL (Sec.~\ref{sec:lumfun_scalefit}). 

\begin{figure}[h!]
\centering

\includegraphics[trim={3.7cm 2.6cm 3.3cm 3.3cm},clip,width=10cm]{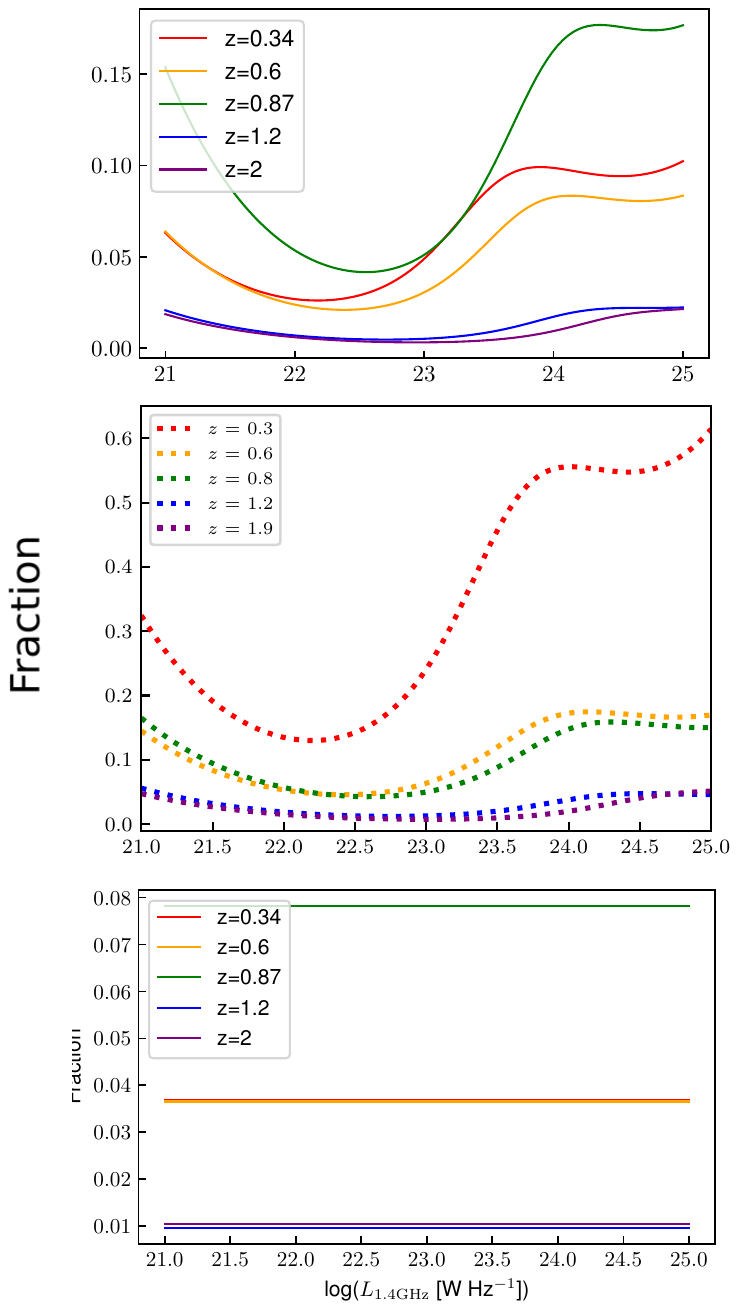}

\caption{Top: Fraction for group galaxies showing their contribution to the total 3 GHz radio luminosity function at different epochs using the power-law, linear regression fit method presented in Sec.~\ref{sec:lumfun_linfit} vs. radio luminosity at 1.4 GHz. Colours represent different redshift bins. Middle: Same as above, but for all massive galaxies ($M_{*} > 10^{11.2} \rm M_{\odot}$) with radio emission at 3 GHz \citep{smolcic17b, laigle16}. 
Bottom: Fraction using the scaled fit method presented in Sec.~\ref{sec:lumfun_scalefit}. A halo mass cut above $10^{13.5}M_{\odot}$ was applied to all plots.}
\label{fig:frac_lum}
\end{figure}

To quantify the contribution of the RLF of group galaxies to the total RLF of the 3 GHz population, we divide the power-law RLF of GGs, assuming a fixed slope of $\gamma=-0.75$, with the total RLF from the 3 GHz sample for each redshift bin. This gives the fractional contribution of group galaxies to the total RLF. In the top panel of Fig.~\ref{fig:frac_lum}, we plot the fraction with respect to the radio luminosity at 1.4 GHz up to $10^{25} \rm ~W ~Hz^{-1}$. The reason for that is the RLF of 3 GHz VLA-COSMOS observations is not well constrained above that luminosity \citep{novak18}. Furthermore, we should consider that our model might be a good representation of the universe at $L > 10^{25} \rm ~W ~Hz^{-1}$, and the COSMOS is not suited for low-$z$ studies due to the small volume coverage at low redshifts. Because of the different total RLF shapes per redshift bin, there is a bump in the curve, as expected. Using a fixed slope of $\gamma=-0.75$ does not impact our calculations as the value is within the errors for the fitted $\gamma$ values presented in Table~\ref{tab:lf_fit_par}. 

We also calculate the RLF of all massive galaxies with $M_{*} > 10^{11.2} \rm M_{\odot}$, which are 3 GHz sources, and compare it to the total RLF. This fraction is shown in the middle panel of Fig.~\ref{fig:frac_lum}. We see an increase in the contribution to the total RLF at $z_{\rm med}$ = 0.3, and at higher redshifts similar to that of GGs. The choice of this stellar mass cut, as a comparison, was motivated by the study of \cite{smolcic17c} in order to select massive galaxies across all redshifts. We will discuss this further in the next section.

The bottom panel of Fig.~\ref{fig:frac_lum} presents fractions using the scaled fit method presented in Sec.~\ref{sec:lumfun_scalefit}. The fraction is the same across the adopted luminosity range, as expected for a scaled fit.

\begin{figure*}
\centering
 \resizebox{\hsize}{!}
 {
 \includegraphics[width=1.08\linewidth]{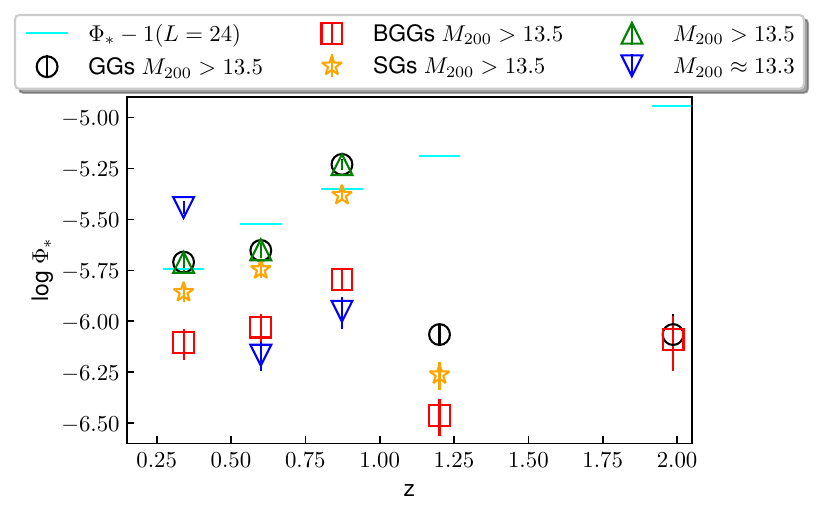}
\includegraphics[width=\linewidth]{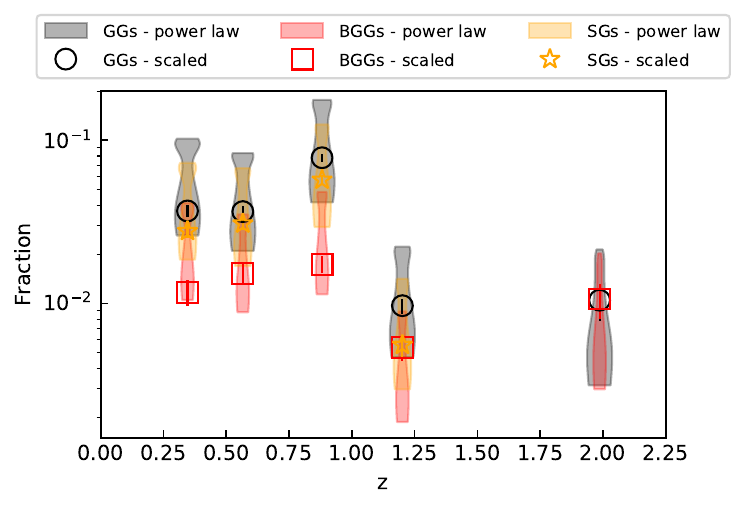}
}

\caption{Left: Radio luminosity function vs. redshift for the group galaxies (black circles), brightest group galaxies BGGs (red squares), and satellite galaxies SGs (yellow stars). Cyan lines show the $\Phi_{*}$ values at log$_{10}(L_{1.4\rm GHz} / \rm W~Hz^{-1})  = 24$ for each redshift bin; these values were placed a dex lower for presentation reasons. The values plotted here are reported in Table~\ref{tab:lf_fit_par} for $\gamma = -0.75$.
Right: The relative contribution of GGs to the total radio luminosity, as described in Sec.~\ref{sec:lumfun_scalefit} for the scaled fit (open symbols) and in Sec.~\ref{sec:lumfun_linfit} for the power-law, linear regression fit (violin plots), vs. redshift. The values in the violin plots show the distribution of fractions along our adopted luminosity range. Black circles denote GGs, red squares denote BGGs, and yellow stars denote SGs. A halo mass cut, log$_{10}(M_{\rm 200c} / \rm M_{\odot}) > 13.5$, was applied. }
\label{fig:rlf_evol}
\end{figure*}

In Fig.~\ref{fig:rlf_evol} we plot the radio luminosity function and the fraction, for the scaled and power-law fits, in relation to redshift for the GGs, BGG, and SGs, and discuss this in the following section.

\section{Evolution of the RLF in galaxy groups}
\label{sec:evolution}

The top panel of Fig.~\ref{fig:frac_lum} shows that the contribution of GGs to the total RLF increases from $z \sim$ 2 to 0.07, and in particular for objects above radio luminosities at 10$^{23}\rm W~Hz^{-1}$. This picture suggests an evolutionary scenario for the RLF of galaxy groups. We investigate this further by plotting in Fig.~\ref{fig:rlf_evol} the RLF of GGs, BGGs, and SGs (left panel), and their relative contribution to the total 3 GHz RLF as a function of redshift (right panel). The GG RLF has a low value at $ \sim$ 2 down to $z \sim$ 1.25 followed by a sharp increase in the GG RLF at $z \sim$ 1 by a factor of 6, and then a smooth decline, mimicking a mild evolution by a factor of 2. 
This is an interesting trend, which is not observed in the total RLF. As seen by the normalisation value of the total $\Phi_{*}$ for a fixed radio luminosity at $10^{24} \rm W~Hz^{-1}$, shown with cyan, galaxies in the 3 GHz sample display a decrease in their RLF with redshift across all redshifts, while the RLF of GGs increases at $z \sim$ 1 (left panel of Fig.~\ref{fig:rlf_evol}); the peak at $z_{\rm med} \sim 0.9$ coincides with known overdensities in the COSMOS field. Interestingly, \cite{smolcic17c} show that a similar trend can be reproduced with galaxies. In their Fig. 1, they present a slight increase in the median values of $M_{*}$ in their radio excess sample up to redshift $z=1$, and depletion of massive galaxies above $z>1$, which we also see in the X-ray groups. We observed this at a median value $\sim10^{11.2} M_{\odot}$. At the middle panel of Fig.~\ref{fig:frac_lum} we see that massive galaxies above $10^{11.2} M_{\odot}$ contribute a large fraction to the total 3 GHz RLF below $z<1$. This suggests that not all massive galaxies are in groups, but those that are, remain radio active (Fig.~\ref{fig:frac_lum}). The GG contribution to the RLF has a nearly flat but slightly enhanced behaviour below $z <$ 0.75, while the GGs RLF does not exhibit a large contribution of radio emission at the lowest redshift bin like the massive galaxies RLF does. This suggests those massive galaxies are either in the field or occupy halo masses below our adopted cut at $log_{10}(M_{\rm 200c} / M_{\odot}) > 13.5$.

The RLF of SGs dominates the RLF of group galaxies up to redshift of $z_{\rm med} \sim$ 1.2, with overdensities below $z=1$ (Fig.~\ref{fig:rlf_evol}--Right). The fraction for the linear regression fit method is a range of values that correspond to the adopted luminosity range (see also Fig.~\ref{fig:frac_lum}), plotted as violin plots. Both fractions follow the mild evolutionary trend we observe on the left panel. \cite{scoville13} studied the large-scale structure (LSS) in COSMOS and also report a statistically significant overdensity at $z =$ 0.93. 
Additionally, the strongest density peaks, where we have massive clusters in COSMOS, are at redshifts 0.37, 0.73, and 0.83. Our $0.4<z<0.7$ bin misses LSS on both ends.
Above $z \sim$ 2 we do not currently have a large-enough number of SGs to perform a robust analysis. This is likely to improve with future observations. 
The relative contribution of the SGs to the RLF of group galaxies is higher by a factor of 2 than that of BGGs below $z \sim$ 1. Additionally, BGGs contribute a  small amount to the RLF of GGs, as seen by the left panel of Fig.~\ref{fig:rlf_evol}, despite being the most massive galaxies of the group. This is a very interesting result highlighting the importance of identifying the member group galaxies within a group and the need for high sensitivity and high-resolution observations.

For reference we have split the redshift bins into low and high halo mass objects. Objects with group masses below $10^{13.5}~M_{\odot}$ contribute significantly to the lowest redshift bins and are linked to SGs, but this contribution is not taken into account in our analysis, to ensure our sample is complete (see Sec.~\ref{sec:data}).
The low halo mass points (Table~\ref{tab:lf_fit_par} \& Fig.~\ref{fig:lfgrid}) show a faster turnover as we do not expect to detect many low mass, high luminosity objects.

\cite{yuan16}, who studied brightest cluster galaxies (BCGs), found that RLFs of 7138 BCGs in the range 0.05 $< z <$ 0.45 do not show significant evolution with redshift. This no-evolution pattern of BCGs agrees with our results for BGGs in COSMOS. At the left panel of Fig.~\ref{fig:rlf_evol}, we see that the RLF of BGGs fluctuates slightly with redshift, but it is the RLF of satellites that drives the redshift evolution.

\cite{novak18} discuss possible biases which could affect the calculations. These include the assumed shape of the radio SED to be a power law and the radio excess criterion to be too conservative and thus excluding low-luminosity AGN from the sample. We refer the reader to their discussion (see their Section 3.4).
Furthermore, \cite{novak18} discuss possible biases that affect the RLF of the high luminosity bin, i.e., bright radio but faint in the near-infrared sources ($K$ = 24.5 mag). 
We have constrained our sample to halo masses above $10^{13.5} M_{\odot}$, to perform an unbiased analysis. Incidentally, after the halo-mass cut, the remaining group galaxies in our sample are brighter than $K$ = 24.5 mag.

In summary, we observe a nearly flat but slightly enhanced behavior of the contribution of X-ray galaxy groups to the 3~GHz RLF up to $z \sim$ 0.75, driven by SGs and AGN (see Sec.~\ref{sec:agn}) in GGs, followed by an increase and then a sharp drop. This agrees with past studies of the COSMOS field, and in particular with the study of \cite{hale2018}, who in their Fig. 10 showed that the AGN bias starts to deviate from values close to $5\times 10^{13}M_{\odot}$ at $z<$ 1 towards $1\times 10^{13}M_{\odot}$ at $z>$1. This explains the sharp drop we observe at $z>$1, since we are probing halo masses $>10^{13.5}M_{\odot}$ (see bottom panel of Fig.~\ref{fig:z_gg}).

\section{The AGN and SFG contribution to the GG RLF}
\label{sec:agn}

The group galaxy population has a mixture of contributions from AGN and SFGs. To explore how much these populations contribute to the GG RLF, we cross-correlate the X-ray galaxy group catalogue with the sample of \cite{vardoulaki21}, which is a value-added catalogue at 3 GHz VLA-COSMOS, and includes 130 FR-type radio sources \citep[FRI, FRII;][ and hybrids FRI/FRII]{fr74} and 1818 jet-less compact radio AGN (COM AGN), as well as 7232 SFGs (see Table~\ref{tab:fr_nums}). Radio AGN in the \cite{smolcic17b} sample were selected on the basis of their radio excess, as mentioned above. This criterion, due to the 3$\sigma$ cut applied, excludes several FR-type radio AGN, which were identified in \cite{vardoulaki21} and classified as radio AGN because they exhibit jets/lobes. SFGs are objects which do not display radio excess.

\begin{table}
\label{tab:fr_nums}
\begin{center}
\caption{The AGN and SFGs inside X-ray galaxy groups. Data from \citep{vardoulaki21}, cross-correlated with the X-ray galaxy group catalogue \citep[][and in prep.]{gozaliasl19}.}
\renewcommand{\arraystretch}{1.5}
\ifonecol \scriptsize \fi
\begin{tabular}[t]{l c c c c c}
\hline\hline
\multicolumn{1}{l}{Number of sources}& \multicolumn{1}{c}{AGN}  & \multicolumn{1}{c}{SFGs}  \\     
\hline\hline
total 3 GHz VLA-COSMOS & 1948 & 7232 \\
same area \& $z$ as X-ray groups & 1038 & 6452 \\
X-ray group members &  138 & 240\\
BGGs & 67 & 47 \\
SGs & 71 & 193\\ 
\hline
\end{tabular}
\end{center}
\end{table}

\begin{figure}
\centering
\includegraphics[width=\linewidth]{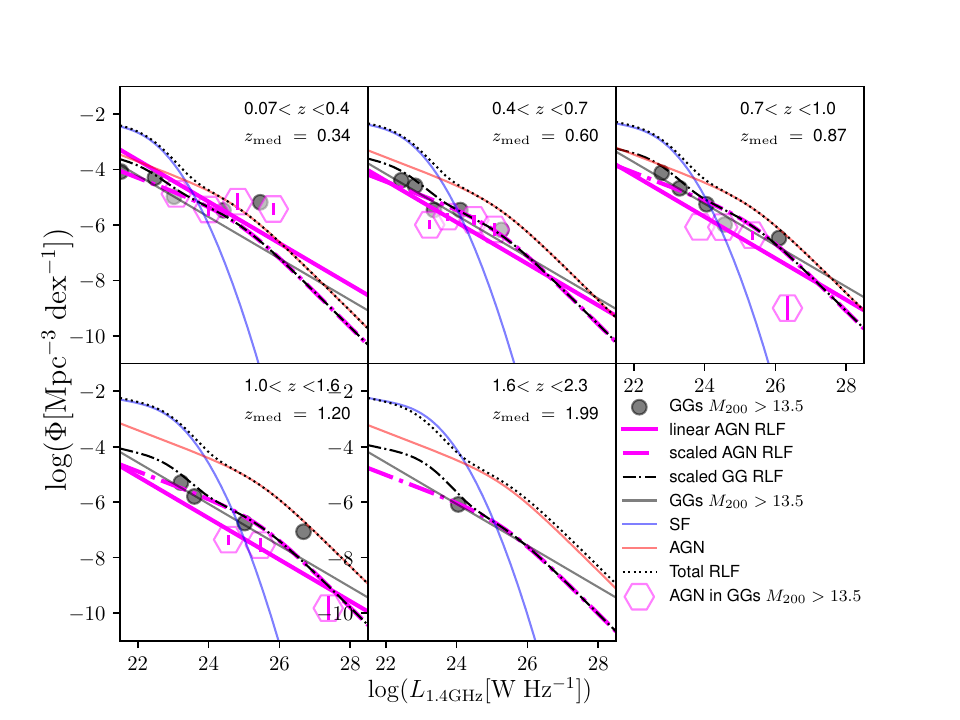}
\includegraphics[width=\linewidth]{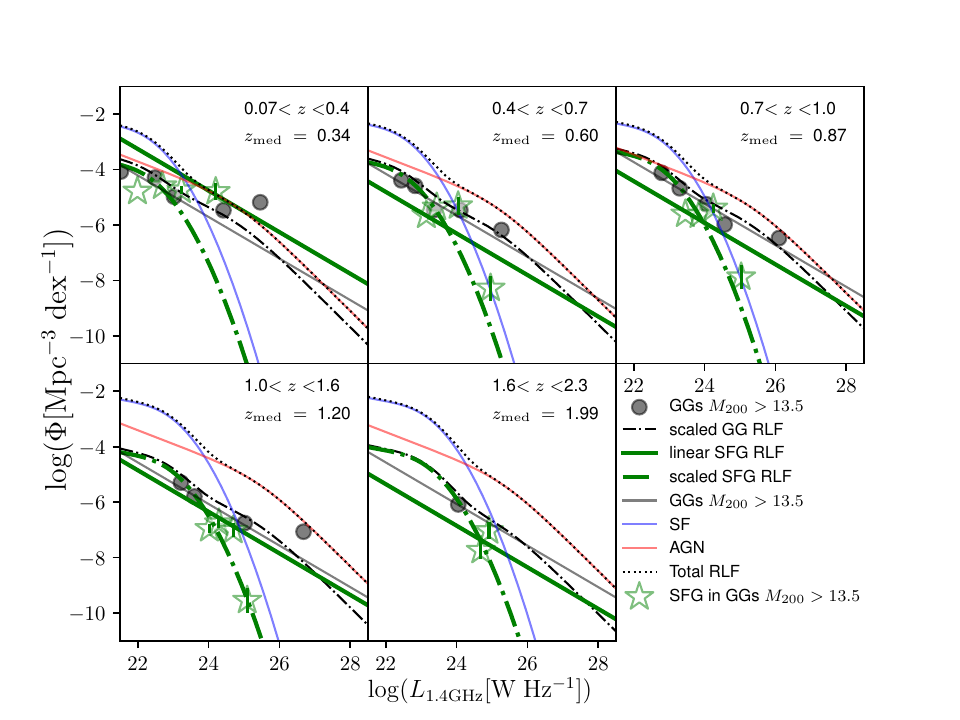}
\caption{Total radio luminosity functions of galaxies in groups, as in Figs~\ref{fig:lfgrid_scale}~\&~\ref{fig:lfgrid}, including RLFs for different populations: radio AGN inside galaxy groups as magenta hexagons (top) and SFGs inside galaxy groups as green stars (bottom). To compare to the GG sample, we normalise the fit to the AGN and SFGs inside groups to $L_{1.4 \rm GHz} = 10^{24}\rm ~W~Hz^{-1}$ and slope of $\gamma$ = -0.75. For comparison, we show the GG sample (black circles for data, and a black solid line with a slope $\gamma$ = -0.75 for the fit). We also plot the scaled AGN and SFG RLF (dashed-dotted lines), as reference. The red solid line shows the RLF for all AGN, the blue solid line for all SFGs, and the dotted black line is the total RLF. A halo mass cut of $M_{\rm 200c}>10^{13.5}M_{\odot}$ was applied. We note, the green solid line at the last bin of the SFG sample is forced to go though the two green stars.}
\label{fig:rlf_agn}
\end{figure}

To quantify the contribution of these populations separately to the group RLF and to the total RLF we calculate their RLF as described in Sec.~\ref{sec:lumfun}, using the $V_{\rm max}$ method. All AGN and SFGs are in groups with halo masses $M_{\rm 200c} > 10^{13.5} \rm M_{\odot}$. 
The results for the AGN and SFG populations inside galaxy groups are shown in Fig.~\ref{fig:rlf_agn}, where we also plot the RLF of AGN and SFGs from the sample of \cite{novak18} as in Fig.~\ref{fig:lfgrid} and the total RLF at 3 GHz. In order to compare the RLF of AGN and SFGs which are GGs and total RLF, we follow the analysis in Sec.~\ref{sec:lumfun_linfit}. We fit a linear regression and normalise it to $10^{24}\rm ~W ~Hz^{-1}$ by applying Eq.~\ref{eq:plaw} for $\gamma$ = -0.75. The results are shown in Fig.~\ref{fig:rlf_agn}. 

For completeness and to enable comparisons between the methods, we also present the scaled AGN and SFG parts of the total RLF to the GG data, and overplot it in Fig.~\ref{fig:rlf_agn}. In Table~\ref{tab:fractions_agn_sfg} we give the scaling coefficients used in Fig.~\ref{fig:lfgrid_scale} at the five redshift bins, separately for the AGN and SFG scaled fits.  Visually, we see that the scaled RLF is not a good fit to the SFG at $z < 0.4$ and to the AGN at $z < 1.6$, inside X-ray galaxy groups. 
We perform a $\chi^{2}$ test for the AGN and SFGs, at all redshift bins and present the results in Table~\ref{tab:chi2_agn_sfg}. The results show that neither of the fitting methods fits the data well.

\begin{table}
\label{tab:fractions_agn_sfg}
\begin{center}
\caption{Scaling coefficients for the functional form of the RLF (scaled fit) for the AGN and SFG populations.}
\renewcommand{\arraystretch}{1.5}
\ifonecol \scriptsize \fi
\begin{tabular}[t]{l c c c c c}
\hline\hline
& \multicolumn{5}{c}{Redshift $z_{\rm med}$}  \\     
& \multicolumn{1}{c}{0.3}& \multicolumn{1}{c}{0.6} & \multicolumn{1}{c}{0.9} & \multicolumn{1}{c}{1.2} & \multicolumn{1}{c}{2} \\ 
\hline\hline
AGN & 0.258& 0.133& 0.224& 0.032& 0.028\\
SFG & 0.041& 0.044& 0.094& 0.012& 0.016\\
\hline
\end{tabular}
\end{center}
\end{table}

\begin{figure}[ht!]
\centering
\includegraphics[width=\linewidth]{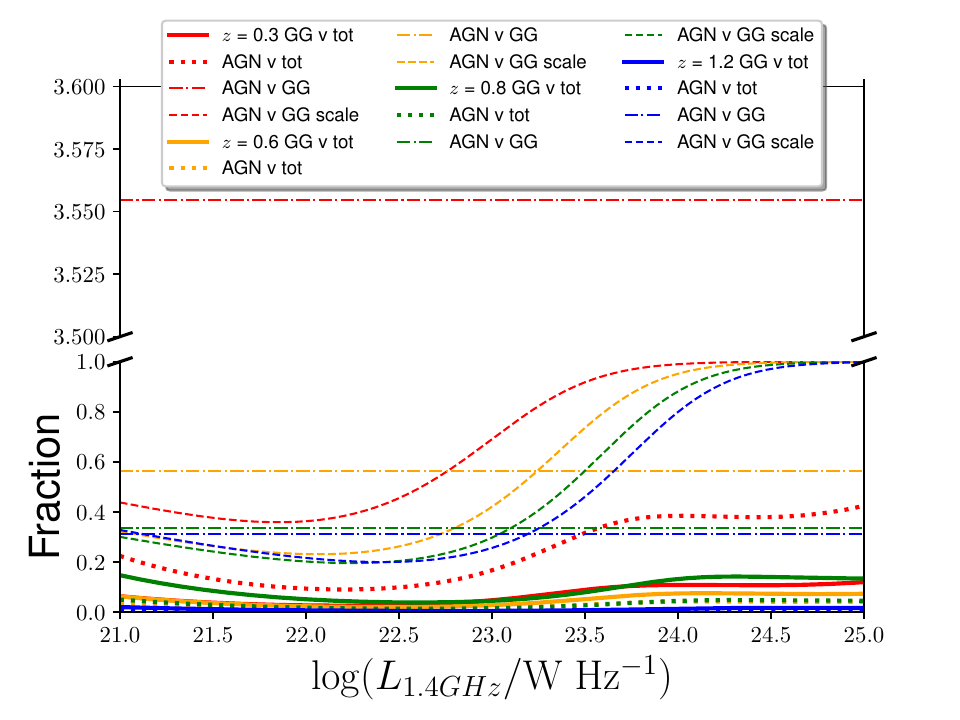}
\includegraphics[width=\linewidth]{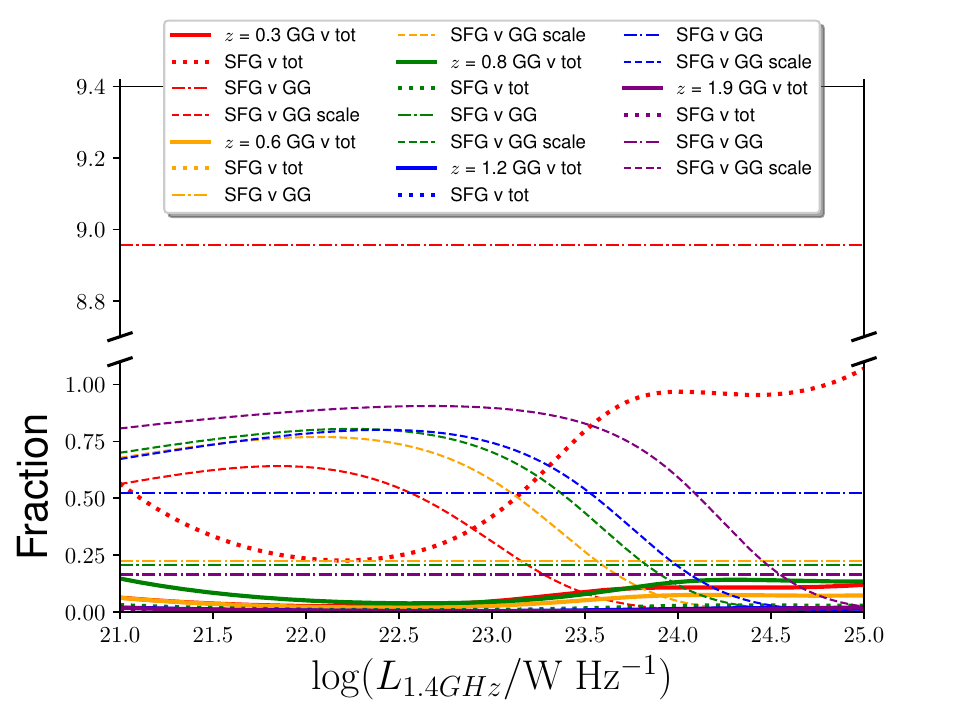}
\caption{Fractional contribution to the total radio luminosity function at different epochs vs radio luminosity at 1.4 GHz for GGs (solid lines as in Fig.~\ref{fig:frac_lum}), and for different populations (dotted lines): AGN (top; labelled AGN v tot) and SFGs (bottom; labelled SFG v tot). Dotted-dashed lines show the fractional contribution of AGN (labelled AGN v GG) and SFG (labelled SFG v GG) to the GG RLF. For reference, we plot with dashed lines the fraction of scaled AGN RLF (top) and scaled SFG RLF (bottom) to scaled GG RLF}. Different colours represent different redshift bins as in Fig.~\ref{fig:frac_lum}. A halo mass cut of $M_{\rm 200c}>10^{13.5}M_{\odot}$ was applied.
\label{fig:fraction_agn}
\end{figure}

We further calculate and plot the fractional contribution of AGN and SFGs which lie inside groups to the total RLF at 3 GHz (Fig.~\ref{fig:fraction_agn}), by replicating Fig.~\ref{fig:frac_lum}. The fraction was calculated by dividing the RLF of AGN and of SFGs inside galaxy groups by the total 3 GHz RLF. The fractions per redshift bin are curved lines due to the total RLF being curved. We find that there is a significant contribution from group AGN and SFGs at redshifts $z <$ 1.6, and very little contribution above. We present the values for these fractions in Table~\ref{tab:agn_fraction}. For completeness, we calculate the fractional contribution of the AGN and SFG RLF to the GG and total RLFs in the case where the scaled-fit method is used. The fractions are also presented in Table~\ref{tab:fr_nums}. In Fig.~\ref{fig:fraction_agn} we overplot the fraction of AGN and SFG RLF to the GG RLF, where both RLFs were calculated using the scaled-fit method. The respective lines follow the scaled AGN and SFG distributions, where AGN contribute more to the total RLF at higher luminosities, while SFGs dominate at lower luminosities. We stress that the scaled method for AGN and SFGs inside galaxy groups in COSMOS, and given the current dataset, is an approximation. It assumes the total 3GHz RLF fits the sub-populations of AGN and SFGs inside galaxy groups. The reason for using a scaled fit is the smaller numbers of objects in galaxy groups compared to the total 3GHz sample. Ideally, with a larger sample of AGN and SFGs inside groups, an MCMC can provide a good fit to the sub-populations. Our analysis suggests that the total 3GHz RLF is not a good fit for individual populations inside galaxy groups and that the picture is more complicated than that.

For the linear regression fit method we get a constant value across all luminosities in Fig.~\ref{fig:fraction_agn} because the divided fits are both linear. The contribution of AGN RLF to the GG RLF is significant at the redshift bin $z_{\rm med}$ = 0.6 of around 56\% and at $z_{\rm med}$ = 0.8 with fraction around 33\%, and dominates the GG RLF. The fraction in SFGs is around 20\% for $z_{\rm med}$ = 0.6 and $z_{\rm med}$ = 0.8, while at $z_{\rm med}$ = 1.2 the SFGs are dominating the GG RLF, with a fraction of 52\%. At $z_{\rm med}$ = 0.3 we also see enhanced contribution in both AGN and SFGs compared to the GG RLF. This can be explained by the linear regression fit being normalised to $10^{24}\rm  ~W ~Hz^{-1}$ and forced to have a slope of $\gamma$ = -0.75. For $z >$ 1.6 the contribution of SFGs to the GG RLF drops sharply and below 1\%, while we do not have AGN above $z >$ 1.6. These findings suggest that both AGN and SFGs contribute to the GG RLF, with the AGN contribution peaking around $z \sim 1$.

There are 67 AGN associated with BGGs and 71 with SGs, as shown in Table~\ref{tab:fr_nums}. For SFGs we get 47 BGGs and 193 SGs. Due to the small number of sources per bin, we cannot replicate Fig.~\ref{fig:rlf_evol} by splitting the AGN and SFGs RLF inside groups in BGGs and SGs and calculating their RLF. Fig.~\ref{fig:rlf_evol} suggests the evolution of the GG RLF is driven by satellites. Based on our results on from Fig.~\ref{fig:fraction_agn}, at the $z_{\rm med}$ = 0.3 redshift bin, the SG AGN or SFGs are responsible for the peak of the GG RLF, while at $z_{\rm med}$ = 0.8 the increase is mainly driven by AGN. 

How much of the AGN contribution to the GG RLF comes from extended radio emission, given the capabilities of the 3 GHz VLA-COSMOS survey, is not easy to estimate due to sample size limitations. From Table~\ref{tab:fr_nums} we see that $\sim$ 82\% of AGN inside galaxy groups are jet-less AGN. But in order to robustly answer this question we need to separate FRs and COM AGN inside groups and calculate their RLFs per redshift bin, as above, which we cannot do given the small number of FRs per redshift bin. 
To get an idea of how extended the FRs within the AGN sample are, we have a look at the linear projected sizes $D$ of FRs in \cite{vardoulaki21}. The sensitivity and resolution of the 3 GHz VLA-COSMOS survey are 2.3 $\mu$Jy/beam and 0".75, respectively. This means that we are able to resolve and disentangle structures of $\sim$ 6 kpc at $z \sim$ 2. The smallest FR reported in \cite{vardoulaki21} has $D$ = 8.1 kpc at $z = $ 2.467, just above the resolution limit, and the smallest edge-brightened FR has $D$ = 24.3 kpc at $z =$ 1.128, where the lobes are separated by 8 kpc. Inside X-ray galaxy groups, the smallest FR has $D$ = 13.37 kpc at $z$ = 0.38 with the most extended having $D$ = 608.4 kpc and $z$ = 1.168; this is also the most extended object in the \cite{vardoulaki21} FR sample. Future surveys with increased sensitivity and resolution will be able to resolve jets and lobes in AGN which appear compact at 3 GHz VLA-COSMOS. With future observations at larger sky areas and improved statistics, we will be in a better position to answer this question.

\cite{nobels22} show via hydrodynamical simulations of galaxy groups/clusters with masses above $M_{\rm 200c} > 10^{13.5} M_{\odot}$, like the ones studied here, a cyclical behaviour of AGN quenching and star formation activity: long periods where star formation is quenched by the AGN are followed by shorter periods of star formation and black hole accretion. This is because the reduction of AGN feedback makes the ICM unstable to precipitation and thus initiating a new episode of intense star formation. Furthermore, \cite{pasini20} report that feedback mechanisms in groups and clusters of galaxies are similar. In our study, we find that the AGN contribution to the galaxy groups RLF dominates at redshifts up to $z_{\rm med}$ = 0.8. The hosts of these AGN at 3 GHz VLA-COSMOS are quenched, based on the study of \cite{vardoulaki21}. AGN at $z_{\rm med}$ = 0.8 show low star-formation rates (SFR$_{\rm med} \sim$ 8 $M_{\odot}$/yr) compared to SFGs at similar redshifts (SFR$_{\rm med} \sim$ 24 $M_{\odot}$/yr). At lower redshifts ($z_{\rm med}$ = 0.3), both AGN and SFGs populations show low median SFRs ($\sim$ 1.2 and $\sim$ 3.4 $M_{\odot}$/yr, respectively). The median SFR in the field shows similar median values compared to the one inside galaxy groups for $z_{\rm med}$ = 0.3 \& 0.8, for both AGN and SFGs. 

To verify the cyclical behaviour presented in \cite{nobels22} a study of the duty cycle of individual objects is needed, which is beyond the scope of this analysis. A thorough analysis comparing AGN and SFGs in relation to large-scale environment is presented in \cite{vardoulaki21}, and we refer the reader to that study. Detailed investigation of AGN feedback since $z \ 5$ at 3GHz COSMOS is presented in the studies of \cite{smolcic17c} and \cite{Ceraj2018}.

\begin{table*}[ht!]
\begin{center}
\caption{Fractional contribution $f$ of the AGN and SFG linear (scaled; at $log_{10}(L_{\rm 1.4~GHz}/\rm W~Hz^{-1}) = 23 \& 25$) RLF inside groups to the linear (scaled) GG RLF, and to the total RLF at $L_{\rm 1.4~GHz}$ =  10$^{25}$ and $L_{\rm 1.4~GHz}$ =  10$^{25}\rm ~W~Hz^{-1}$, as in Fig.~\ref{fig:fraction_agn}.}
\label{tab:agn_fraction}
\renewcommand{\arraystretch}{1.5}
\ifonecol \scriptsize \fi
\begin{tabular}[t]{l | c c c | c c c }
\hline\hline
\multicolumn{1}{l}{$z_{\rm med}$} & \multicolumn{1}{l}{$f_{\rm AGN-GG}$} & \multicolumn{1}{l}{$f_{\rm AGN-tot23}$} & \multicolumn{1}{l}{$f_{\rm AGN-tot25}$} & \multicolumn{1}{l}{$f_{\rm SFG-GG}$} & \multicolumn{1}{l}{$f_{\rm SFG-tot23}$} & \multicolumn{1}{l}{$f_{\rm SFG-tot25}$}\\     
\hline\hline
0.3 & 3.55 (0.69; 0.99) & 0.16 (0.06)   & 0.42 (0.25) & 8.95 (0.30; 2$\times10^{-5}$) & 0.41 (0.03) & 1.06 (7$\times10^{-6}$)\\
0.6 & 0.56 (0.42; 0.99) & 0.01 (0.02)   & 0.04 (0.13) & 0.22 (0.57; 1$\times10^{-4}$) & 0.01 (0.03) & 0.01 (2$\times10^{-5}$)\\
0.8 & 0.33 (0.29; 0.99) & 0.01 (0.03)   & 0.04 (0.33) & 0.20 (0.70; 7$\times10^{-4}$) & 0.01 (0.07) & 0.03 (1$\times10^{-4}$)\\
1.2 & 0.31 (0.25; 0.99) & 0.001 (0.003) & 0.01 (0.03) & 0.52 (0.74; 6$\times10^{-5}$) & 0.002 (0.01) & 0.01 (6$\times10^{-5}$)\\
1.9 & $-$         & $-$           & $-$         & 0.16 (0.89; 0.02) & 4$\times10^{-4}$ (0.01) & 0.003 (6$\times10^{-4}$)\\
\hline
\end{tabular}
\end{center}
\end{table*}

Our analysis suggests that the bulk of high-$z$ $log_{10}(M_{\rm 200c} / \rm M_{\odot}) > 13.5$ groups must have been forming recently, and so the cooling has not been established. This is linked to the drop in occurrence of AGN in groups at high $z$ by a factor of 6, suggesting that AGN feedback is lower by a factor of 6 at high redshifts. Hence, AGN feedback in the groups we are studying ($log_{10}(M_{\rm 200c}/\rm M_{\odot})$ in the range 13.5-14.5) must be a recent phenomenon. There seems to be a change in the way groups operate above $z > 1$, with a faster evolution. Mass changes quickly and there is not enough time to virialise. Due to the lack of virialisation, the cooling does not start and the AGN activity is suppressed. This change can be triggered by 1) high thermalisation of matter, which is not sufficient in this case; 2) dynamically young groups where gas cooling does not happen. On the other hand, low-mass groups form at $z = 6$. These are found to host radio AGN and have time to virialise, cool, and provide feedback. Additionally, cooling times for energetic electrons is much lower at high-$z$.

\section{Summary and Conclusions}
\label{sec:summary}
We presented a study of radio luminosity functions, RLFs, of group galaxies in the COSMOS field, based on data from the VLA-COSMOS 3~GHz Large Project \citep{smolcic17a} and the X-ray galaxy groups catalogue \citep[][and in prep.]{gozaliasl19}. 
The X-ray galaxy groups cover halo masses in the range $M_{\rm 200c} = 8 \times 10^{12} - 3 \times 10^{14} M_{\odot}$ and the redshift range 0.07 $< z <$ 2.3. To probe the same group population at all redshifts, we applied a halo-mass cut and only selected groups with halo masses $M_{\rm 200c} > 10^{13.5} M_{\odot}$. Furthermore, we applied completeness corrections to the calculation of the RLF \citep{novak17} and all galaxy-group members are brighter than $K$ = 24.5 mag, which allows for an unbiased analysis.

We calculated the RLF of group galaxies based on the $V_{\rm max}$ method and compared it to the 3 GHz RLF from \cite{novak18} who fitted the total RLF with pure luminosity evolution models that depend on redshift. The AGN and SFG populations, characterised by the radio excess parameter, were fitted with a Markov chain Monte Carlo algorithm. We fitted the group galaxies' (GGs) RLFs using two methods, a) by scaling the total 3 GHz RLF, and b) with a linear (power-law) fit, and estimated their contribution to the total RLF. We also studied how much satellites (SGs), brightest group galaxies (BGGs), AGN, and SFGs contribute to the RLF of galaxy groups and to the total 3 GHz RLF. The two fitting methods provide similar results, with the ratio between data and model (Fig.~\ref{fig:residuals}) suggesting the power-law fit proving slightly better for the GG RLF. The linear regression fit is the adopted method for the interpretation of the results.

Our main results are summarised below:

\begin{enumerate}
    \item The relative contribution of the group galaxies to the total 3 GHz radio luminosity function in galaxies in the COSMOS field generally decreases with increasing redshift, from 4\% at low $z$, to 1\% at $z >$ 1, with an overdensity below $z <$ 1, in line with large-scale structure studies of the COSMOS field.
    \item The GG RLF has a low value at $ \sim$ 2 down to  $z \sim$ 1.25 followed by a sharp increase in the GG RLF at $z \sim$ 1 by a factor of 6, and then a smooth decline, which is driven mainly by satellite GGs. The latter suggests a mild evolution in the RLF of GGs from $z \sim$ 1 to 0.07 by a factor of 3.
    \item The RLF of SGs dominates the RLF of group galaxies up to redshift of $z \sim$ 1.2, where we observe a drop in the RLF of both BGGs and SGs. 
    \item The AGN dominate the GG RLF at $z \sim$ 1, while the SFGs dominate the GG RLF at $z_{\rm med}$ = 1.2. 
\end{enumerate}

In summary, we observed a nearly flat but enhanced behavior of the contribution of galaxy groups to the total 3~GHz RLF up to $z \sim$ 0.75, driven by SGs and AGN in GGs, followed by an increase and then a sharp drop, which agrees with the literature and is related to AGN occupying less massive halos above $z >$ 1. The enhanced contribution and sharp drop are not driven by a possible sensitivity drop at high redshifts, but the actual abundance of massive groups, which is enhanced in high-density peaks with regards to normal galaxies, and which creates an enhancement of the fractional contribution of radio galaxies. In the case where all the galaxies would be groups of similar mass, but we would detect only some with X-rays, the ratio would have stayed the same, independent of the density of the field.

Another important result of this analysis is the RLF for group galaxies itself, as well as the contribution of the satellites and BGGs in group environments, which is a major observational constraint for tuning the models. Our study provides an observational probe for the accuracy of the numerical predictions of the radio emission in galaxies in a group environment. Finally, our results show a drop in occurrence of AGN in groups at high $z$ by a factor of 6, suggesting that AGN feedback is lower by a factor of 6 at high redshifts. The bulk of high-$z$ $log_{10}(M_{\rm 200c}/\rm M_{\odot}) > 13.5$ groups must have been forming recently, and so the cooling has not been established. AGN at high-$z$ occupy low halo mass systems ($\approx10^{13.3}\rm~ M_{\odot}$), revealing the details on the processes accountable for the galaxy evolution in massive environments.

\begin{acknowledgements}
We would like to thank Aritra Basu for useful discussions. We would like to thank the anonymous referee for a constructive report which significantly improved our manuscript. EV acknowledges support from Carl Zeiss Stiftung with the project code KODAR. 
\end{acknowledgements}

\bibliography{rlf_groups}
\bibliographystyle{aa}

\begin{appendix}
\onecolumn

\section{Numerical results from the calculation of the radio luminosity function in X-ray galaxy groups in COSMOS}

\begin{sidewaystable*}
\caption{Radio properties of Group Galaxies in COSMOS}     
\label{tab:radioprops}      
\centering          
\begin{tabular}{l l l l c c c c c c c c c l}     
\hline\hline
\multicolumn{2}{c}{ID}  & \multicolumn{2}{c}{radio} & \multicolumn{2}{c}{X-ray} & \multicolumn{2}{c}{Redshift $z$} & \multicolumn{1}{c}{$\rm log_{10}($}& \multicolumn{1}{c}{$\rm log_{10}($}& \multicolumn{1}{c}{$log_{10}($}& \multicolumn{1}{c}{}& \multicolumn{1}{c}{BGG} & \multicolumn{1}{c}{Radio}\\ 
\multicolumn{1}{c}{3~GHz} & \multicolumn{1}{c}{X-ray} & \multicolumn{1}{c}{R.A.} & \multicolumn{1}{c}{Dec.} & \multicolumn{1}{c}{R.A.} & \multicolumn{1}{c}{Dec.} & \multicolumn{1}{c}{Radio} & \multicolumn{1}{c}{X-ray} & \multicolumn{1}{c}{$L_{\rm 1.4~GHz}/$} & \multicolumn{1}{c}{$L_{\rm X}/$} & \multicolumn{1}{c}{$M_{200c}/$} & \multicolumn{1}{c}{$R_{200c}$} & \multicolumn{1}{c}{Rank} & \multicolumn{1}{c}{Class} \\
 & & \multicolumn{4}{c}{(deg., J2000.0)} & & & \multicolumn{1}{c}{$\rm W~Hz^{-1})$} & \multicolumn{1}{c}{$10^{42} \rm erg~s^{-1})$} & \multicolumn{1}{c}{$10^{13} \rm M_{\odot})$} & \multicolumn{1}{c}{(deg.)} \\
 \multicolumn{1}{c}{(1)} & \multicolumn{1}{c}{(2)}  & \multicolumn{1}{c}{(3)} & \multicolumn{1}{c}{(4)}& \multicolumn{1}{c}{(5)}& \multicolumn{1}{c}{(6)} & \multicolumn{1}{c}{(7)} & \multicolumn{1}{c}{(8)} & \multicolumn{1}{c}{(9)} & \multicolumn{1}{c}{(10)} & \multicolumn{1}{c}{(11)} & \multicolumn{1}{c}{(12)} & \multicolumn{1}{c}{(13)} & \multicolumn{1}{c}{(14)}\\
\hline\hline
29  &  79  &  150.447297  &  2.053936  &  150.4456  &  2.054150  &  0.323  &  0.324$^{ s }$  &  24.36  &  1.81 $\pm$ 0.86  &  1.73 $\pm$ 0.58  &  0.0417  &  0  & AGN \\ 
35  &  333  &  150.042302  &  2.694820  &  150.04736  &  2.693170  &  0.218  &  0.219$^{ s }$  &  23.50  &  1.03 $\pm$ 0.40  &  1.26 $\pm$ 0.43  &  0.0400  &  0  & AGN \\ 
42  &  11  &  150.185632  &  1.661683  &  150.19728  &  1.658950  &  0.221  &  0.220$^{ s }$  &  23.34  &  2.54 $\pm$ 0.75  &  2.23 $\pm$ 0.24  &  0.0838  &  5  & AGN \\ 
44  &  237  &  150.110399  &  2.708261  &  150.11756  &  2.692520  &  0.351  &  0.349$^{ s }$  &  24.48  &  2.19 $\pm$ 0.95  &  1.96 $\pm$ 0.52  &  0.0468  &  2  & AGN \\ 
45  &  189  &  149.905915  &  2.396440  &  149.91816  &  2.388030  &  0.742  &  0.736$^{ s }$  &  24.40  &  1.82 $\pm$ 1.30  &  1.58 $\pm$ 0.85  &  0.0209  &  3  & AGN \\ 
61  &  134  &  149.649621  &  2.209249  &  149.6507  &  2.211630  &  0.954  &  0.951$^{ s }$  &  23.85  &  2.27 $\pm$ 1.58  &  1.78 $\pm$ 0.88  &  0.0202  &  0  & AGN \\ 
80  &  264  &  149.998389  &  2.769141  &  149.99834  &  2.774410  &  0.166  &  0.165$^{ s }$  &  23.54  &  1.03 $\pm$ 0.45  &  1.28 $\pm$ 0.48  &  0.0516  &  0  & AGN \\ 
88  &  52  &  150.44703  &  1.882829  &  150.44759  &  1.883190  &  0.670  &  0.672$^{ s }$  &  24.02  &  1.84 $\pm$ 1.17  &  1.62 $\pm$ 0.74  &  0.0223  &  0  & AGN \\ 
99  &  25  &  149.884002  &  1.750996  &  149.85402  &  1.770230  &  0.122  &  0.124$^{ s }$  &  22.10  &  1.58 $\pm$ 0.08  &  1.65 $\pm$ -0.05  &  0.0884  &  0  & AGN \\ 
103  &  143  &  150.207423  &  2.281893  &  150.22014  &  2.278510  &  0.917  &  0.885$^{ s }$  &  24.53  &  2.18 $\pm$ 1.43  &  1.75 $\pm$ 0.79  &  0.0207  &  3  & AGN \\ 
119  &  245  &  150.695821  &  2.760741  &  150.69144  &  2.754350  &  0.630  &  0.646$^{ s }$  &  23.36  &  2.05 $\pm$ 1.34  &  1.76 $\pm$ 0.84  &  0.0257  &  2  & AGN \\ 
127  &  187  &  149.623522  &  2.399203  &  149.62376  &  2.399380  &  0.846  &  0.840$^{ s }$  &  24.10  &  2.33 $\pm$ 1.60  &  1.86 $\pm$ 0.92  &  0.0231  &  0  & AGN \\ 
149  &  44  &  149.805372  &  1.874522  &  149.79852  &  1.872330  &  1.329  &  1.350$^{ p }$  &  25.10  &  2.40 $\pm$ 1.98  &  1.71 $\pm$ 1.07  &  0.0155  &  0  & AGN \\ 
156  &  217  &  150.007104  &  2.453454  &  150.0037  &  2.454870  &  0.731  &  0.732$^{ s }$  &  24.14  &  1.73 $\pm$ 1.15  &  1.52 $\pm$ 0.73  &  0.0196  &  0  & AGN \\ 
178  &  46  &  150.250884  &  1.863985  &  150.24614  &  1.865000  &  0.530  &  0.529$^{ s }$  &  23.81  &  1.61 $\pm$ 1.01  &  1.52 $\pm$ 0.71  &  0.0246  &  0  & AGN \\ 
182  &  190  &  149.441577  &  2.425937  &  149.45692  &  2.433530  &  0.481  &  0.484$^{ s }$  &  24.03  &  2.16 $\pm$ 1.36  &  1.89 $\pm$ 0.89  &  0.0347  &  0  & AGN \\ 
183  &  149  &  150.415664  &  2.430186  &  150.42235  &  2.428000  &  0.124  &  0.124$^{ s }$  &  22.34  &  1.81 $\pm$ 0.16  &  1.79 $\pm$ -0.06  &  0.0986  &  0  & AGN \\ 
187  &  311  &  149.942948  &  2.600608  &  149.93887  &  2.606860  &  0.344  &  0.342$^{ s }$  &  24.84  &  1.50 $\pm$ 0.56  &  1.52 $\pm$ 0.38  &  0.0339  &  2  & AGN \\ 
203  &  291  &  149.478815  &  2.521843  &  149.48956  &  2.516580  &  0.092  &  0.092$^{ s }$  &  22.10  &  0.83 $\pm$ 0.12  &  1.17 $\pm$ 0.26  &  0.0806  &  84  & SFG \\ 
211  &  215  &  150.172574  &  2.523342  &  150.17097  &  2.523630  &  0.697  &  0.697$^{ s }$  &  24.14  &  1.58 $\pm$ 1.13  &  1.44 $\pm$ 0.77  &  0.0190  &  1  & AGN \\ 
213  &  24  &  150.30117  &  1.689921  &  150.29169  &  1.689350  &  0.530  &  0.527$^{ s }$  &  24.14  &  1.88 $\pm$ 1.02  &  1.70 $\pm$ 0.63  &  0.0281  &  1  & AGN \\ 
220  &  203  &  149.666143  &  2.474287  &  149.65762  &  2.479720  &  0.956  &  0.948$^{ s }$  &  23.50  &  2.28 $\pm$ 1.59  &  1.78 $\pm$ 0.89  &  0.0201  &  2  & SFG \\ 
226  &  36  &  149.894896  &  1.764946  &  149.89491  &  1.764970  &  0.530  &  0.531$^{ s }$  &  23.34  &  1.76 $\pm$ 1.09  &  1.61 $\pm$ 0.74  &  0.0262  &  1  & AGN \\ 
235  &  20  &  150.325431  &  1.607013  &  150.32584  &  1.605100  &  0.222  &  0.227$^{ s }$  &  23.01  &  1.13 $\pm$ 0.43  &  1.33 $\pm$ 0.42  &  0.0419  &  8  & AGN \\ 
246  &  126  &  150.750943  &  2.199992  &  150.74525  &  2.196520  &  0.372  &  0.372$^{ s }$  &  23.93  &  1.68 $\pm$ 0.99  &  1.62 $\pm$ 0.72  &  0.0345  &  0  & SFG \\ 
251  &  173  &  150.058047  &  2.380425  &  150.05441  &  2.374090  &  0.347  &  0.347$^{ s }$  &  22.63  &  1.37 $\pm$ 0.66  &  1.44 $\pm$ 0.51  &  0.0314  &  1  & AGN \\ 
261  &  271  &  149.91793  &  2.701673  &  149.91853  &  2.700660  &  0.891  &  0.890$^{ s }$  &  24.22  &  2.33 $\pm$ 1.60  &  1.84 $\pm$ 0.90  &  0.0221  &  4  & AGN \\ 
278  &  302  &  150.090688  &  2.205629  &  150.09077  &  2.205680  &  0.426  &  0.427$^{ s }$  &  23.37  &  1.01 $\pm$ 0.63  &  1.18 $\pm$ 0.57  &  0.0220  &  0  & AGN \\ 
279  &  221  &  150.565828  &  2.496444  &  150.57024  &  2.498640  &  1.146  &  1.147$^{ s }$  &  23.58  &  2.42 $\pm$ 2.01  &  1.80 $\pm$ 1.17  &  0.0183  &  0  & AGN \\ 
302  &  14  &  150.247725  &  1.542408  &  150.24973  &  1.543310  &  0.886  &  0.886$^{ p }$  &  24.21  &  2.50 $\pm$ 1.88  &  1.88 $\pm$ 1.06  &  0.0204  &  0  & AGN \\ 

\\
\hline
\end{tabular}
\tablefoot{Basic radio properties of group galaxies in COSMOS. \textbf{Columns 1 \& 2}: The 3~GHz radio ID \citep{smolcic17a} and the X-ray galaxy group ID \citep[][and in prep.]{gozaliasl19}, respectively. \textbf{Columns 3 \& 4:} Right Ascension (R.A.) and Declination (Dec.) of the radio position in degrees, respectively. \textbf{Columns 5 \& 6:} Right Ascension (R.A.) and Declination (Dec.) of the X-ray galaxy group position in degrees, respectively. \textbf{Columns 7 \& 8:} Redshift of the radio source and the X-ray galaxy group, respectively. The character 's' denotes spectroscopic and 'p' denotes photometric redshift for the X-ray groups. \textbf{Column 9:} Radio luminosity at 1.4~GHz in W~Hz$^{-1}$. \textbf{Column 10:} X-ray galaxy group luminosity and error in $10^{42} \rm erg~s^{-1}$. \textbf{Column 11:} Halo mass at $M_{\rm 200c}$ in $10^{13} \rm M_{\odot}$ and error. \textbf{Column 12:} Virial radius of the group $R_{200c}$ in degrees.  \textbf{Column 13:} BBG rank, where 0 denotes if the group galaxy is the brightest of the X-ray galaxy group and values $>$ 0 are for satellite galaxies. \textbf{Column 14:} Radio classification based on a combination of the radio excess parameter \citep{smolcic17b, delvecchio17} and objects having radio jets \citep{vardoulaki21};  'AGN' exhibit radio excess or have radio jets, while 'SFG' do not. 
} 

\end{sidewaystable*}

\addtocounter{table}{-1}

\begin{sidewaystable*}
\caption{Radio properties of Group Galaxies in COSMOS}     
\label{tab:radioprops}      
\centering          
\begin{tabular}{l l l l c c c c c c c c c l}     
\hline\hline
\multicolumn{2}{c}{ID}  & \multicolumn{2}{c}{radio} & \multicolumn{2}{c}{X-ray} & \multicolumn{2}{c}{Redshift $z$} & \multicolumn{1}{c}{$\rm log_{10}($}& \multicolumn{1}{c}{$\rm log_{10}($}& \multicolumn{1}{c}{$log_{10}($}& \multicolumn{1}{c}{}& \multicolumn{1}{c}{BGG} & \multicolumn{1}{c}{Radio}\\ 
\multicolumn{1}{c}{3~GHz} & \multicolumn{1}{c}{X-ray} & \multicolumn{1}{c}{R.A.} & \multicolumn{1}{c}{Dec.} & \multicolumn{1}{c}{R.A.} & \multicolumn{1}{c}{Dec.} & \multicolumn{1}{c}{Radio} & \multicolumn{1}{c}{X-ray} & \multicolumn{1}{c}{$L_{\rm 1.4~GHz}/$} & \multicolumn{1}{c}{$L_{\rm X}/$} & \multicolumn{1}{c}{$M_{200c}/$} & \multicolumn{1}{c}{$R_{200c}$} & \multicolumn{1}{c}{Rank} & \multicolumn{1}{c}{Class} \\
 & & \multicolumn{4}{c}{(deg., J2000.0)} & & & \multicolumn{1}{c}{$\rm W~Hz^{-1})$} & \multicolumn{1}{c}{$10^{42} \rm erg~s^{-1})$} & \multicolumn{1}{c}{$10^{13} \rm M_{\odot})$} & \multicolumn{1}{c}{(deg.)} \\
 \multicolumn{1}{c}{(1)} & \multicolumn{1}{c}{(2)}  & \multicolumn{1}{c}{(3)} & \multicolumn{1}{c}{(4)}& \multicolumn{1}{c}{(5)}& \multicolumn{1}{c}{(6)} & \multicolumn{1}{c}{(7)} & \multicolumn{1}{c}{(8)} & \multicolumn{1}{c}{(9)} & \multicolumn{1}{c}{(10)} & \multicolumn{1}{c}{(11)} & \multicolumn{1}{c}{(12)} & \multicolumn{1}{c}{(13)} & \multicolumn{1}{c}{(14)}\\
\hline\hline
309  &  10  &  150.721624  &  1.526961  &  150.75047  &  1.531030  &  0.680  &  0.669 $^{ s }$  &  24.62  &  2.64 $\pm$ 2.09  &  2.13 $\pm$ 1.36  &  0.0332  &  38  & SFG \\ 
315  &  17  &  149.964109  &  1.680330  &  149.96375  &  1.680220  &  0.372  &  0.373 $^{ s }$  &  23.19  &  2.24 $\pm$ 1.08  &  1.98 $\pm$ 0.62  &  0.0453  &  0  & AGN \\ 
316  &  88  &  150.502395  &  2.068242  &  150.49783  &  2.069730  &  0.438  &  0.438 $^{ s }$  &  23.14  &  1.69 $\pm$ 0.81  &  1.61 $\pm$ 0.52  &  0.0299  &  0  & AGN \\
\hline
\end{tabular}
\tablefoot{(Continued)}
\end{sidewaystable*}

\addtocounter{table}{-1}

\begin{sidewaystable*}
\caption{Radio properties of Group Galaxies in COSMOS}     
\label{tab:radioprops}      
\centering          
\begin{tabular}{l l l l c c c c c c c c c l}     
\hline\hline
\multicolumn{2}{c}{ID}  & \multicolumn{2}{c}{radio} & \multicolumn{2}{c}{X-ray} & \multicolumn{2}{c}{Redshift $z$} & \multicolumn{1}{c}{$\rm log_{10}($}& \multicolumn{1}{c}{$\rm log_{10}($}& \multicolumn{1}{c}{$log_{10}($}& \multicolumn{1}{c}{}& \multicolumn{1}{c}{BGG} & \multicolumn{1}{c}{Radio}\\ 
\multicolumn{1}{c}{3~GHz} & \multicolumn{1}{c}{X-ray} & \multicolumn{1}{c}{R.A.} & \multicolumn{1}{c}{Dec.} & \multicolumn{1}{c}{R.A.} & \multicolumn{1}{c}{Dec.} & \multicolumn{1}{c}{Radio} & \multicolumn{1}{c}{X-ray} & \multicolumn{1}{c}{$L_{\rm 1.4~GHz}/$} & \multicolumn{1}{c}{$L_{\rm X}/$} & \multicolumn{1}{c}{$M_{200c}/$} & \multicolumn{1}{c}{$R_{200c}$} & \multicolumn{1}{c}{Rank} & \multicolumn{1}{c}{Class} \\
 & & \multicolumn{4}{c}{(deg., J2000.0)} & & & \multicolumn{1}{c}{$\rm W~Hz^{-1})$} & \multicolumn{1}{c}{$10^{42} \rm erg~s^{-1})$} & \multicolumn{1}{c}{$10^{13} \rm M_{\odot})$} & \multicolumn{1}{c}{(deg.)} \\
 \multicolumn{1}{c}{(1)} & \multicolumn{1}{c}{(2)}  & \multicolumn{1}{c}{(3)} & \multicolumn{1}{c}{(4)}& \multicolumn{1}{c}{(5)}& \multicolumn{1}{c}{(6)} & \multicolumn{1}{c}{(7)} & \multicolumn{1}{c}{(8)} & \multicolumn{1}{c}{(9)} & \multicolumn{1}{c}{(10)} & \multicolumn{1}{c}{(11)} & \multicolumn{1}{c}{(12)} & \multicolumn{1}{c}{(13)} & \multicolumn{1}{c}{(14)}\\
\hline\hline
832  &  63  &  150.360853  &  1.931518  &  150.35324  &  1.933030  &  1.234  &  1.233$^{ p }$  &  23.77  &  2.38 $\pm$ 1.79  &  1.74 $\pm$ 0.94  &  0.0167  &  1  & AGN \\ 
846  &  282  &  150.359573  &  2.659472  &  150.35358  &  2.643340  &  0.397  &  0.396$^{ s }$  &  22.27  &  1.35 $\pm$ 0.65  &  1.41 $\pm$ 0.49  &  0.0277  &  1  & AGN \\ 
879  &  322  &  150.228073  &  2.269746  &  150.2254  &  2.268730  &  0.678  &  0.677$^{ p }$  &  23.37  &  1.65 $\pm$ 1.16  &  1.49 $\pm$ 0.78  &  0.0202  &  1  & SFG \\
\hline
\end{tabular}
\tablefoot{(Continued)}
\end{sidewaystable*}

\addtocounter{table}{-1}

\begin{sidewaystable*}
\caption{Radio properties of Group Galaxies in COSMOS}     
\label{tab:radioprops}      
\centering          
\begin{tabular}{l l l l c c c c c c c c c l}     
\hline\hline
\multicolumn{2}{c}{ID}  & \multicolumn{2}{c}{radio} & \multicolumn{2}{c}{X-ray} & \multicolumn{2}{c}{Redshift $z$} & \multicolumn{1}{c}{$\rm log_{10}($}& \multicolumn{1}{c}{$\rm log_{10}($}& \multicolumn{1}{c}{$log_{10}($}& \multicolumn{1}{c}{}& \multicolumn{1}{c}{BGG} & \multicolumn{1}{c}{Radio}\\ 
\multicolumn{1}{c}{3~GHz} & \multicolumn{1}{c}{X-ray} & \multicolumn{1}{c}{R.A.} & \multicolumn{1}{c}{Dec.} & \multicolumn{1}{c}{R.A.} & \multicolumn{1}{c}{Dec.} & \multicolumn{1}{c}{Radio} & \multicolumn{1}{c}{X-ray} & \multicolumn{1}{c}{$L_{\rm 1.4~GHz}/$} & \multicolumn{1}{c}{$L_{\rm X}/$} & \multicolumn{1}{c}{$M_{200c}/$} & \multicolumn{1}{c}{$R_{200c}$} & \multicolumn{1}{c}{Rank} & \multicolumn{1}{c}{Class} \\
 & & \multicolumn{4}{c}{(deg., J2000.0)} & & & \multicolumn{1}{c}{$\rm W~Hz^{-1})$} & \multicolumn{1}{c}{$10^{42} \rm erg~s^{-1})$} & \multicolumn{1}{c}{$10^{13} \rm M_{\odot})$} & \multicolumn{1}{c}{(deg.)} \\
 \multicolumn{1}{c}{(1)} & \multicolumn{1}{c}{(2)}  & \multicolumn{1}{c}{(3)} & \multicolumn{1}{c}{(4)}& \multicolumn{1}{c}{(5)}& \multicolumn{1}{c}{(6)} & \multicolumn{1}{c}{(7)} & \multicolumn{1}{c}{(8)} & \multicolumn{1}{c}{(9)} & \multicolumn{1}{c}{(10)} & \multicolumn{1}{c}{(11)} & \multicolumn{1}{c}{(12)} & \multicolumn{1}{c}{(13)} & \multicolumn{1}{c}{(14)}\\
\hline\hline
1495  &  254  &  150.10643  &  2.794748  &  150.0965  &  2.806100  &  0.729  &  0.701$^{ s }$  &  23.39  &  1.99 $\pm$ 1.46  &  1.70 $\pm$ 0.96  &  0.0232  &  1  & AGN \\ 
1503  &  88  &  150.484941  &  2.082487  &  150.49783  &  2.069730  &  0.428  &  0.438$^{ s }$  &  22.78  &  1.69 $\pm$ 0.81  &  1.61 $\pm$ 0.52  &  0.0299  &  12  & SFG \\ 
1535  &  241  &  150.490648  &  2.747902  &  150.4879  &  2.751180  &  0.612  &  0.611$^{ s }$  &  23.03  &  1.82 $\pm$ 1.21  &  1.62 $\pm$ 0.80  &  0.0240  &  0  & AGN \\
\hline
\end{tabular}
\tablefoot{(Continued)}
\end{sidewaystable*}

\addtocounter{table}{-1}

\begin{sidewaystable*}
\caption{Radio properties of Group Galaxies in COSMOS}     
\label{tab:radioprops}      
\centering          
\begin{tabular}{l l l l c c c c c c c c c l}     
\hline\hline
\multicolumn{2}{c}{ID}  & \multicolumn{2}{c}{radio} & \multicolumn{2}{c}{X-ray} & \multicolumn{2}{c}{Redshift $z$} & \multicolumn{1}{c}{$\rm log_{10}($}& \multicolumn{1}{c}{$\rm log_{10}($}& \multicolumn{1}{c}{$log_{10}($}& \multicolumn{1}{c}{}& \multicolumn{1}{c}{BGG} & \multicolumn{1}{c}{Radio}\\ 
\multicolumn{1}{c}{3~GHz} & \multicolumn{1}{c}{X-ray} & \multicolumn{1}{c}{R.A.} & \multicolumn{1}{c}{Dec.} & \multicolumn{1}{c}{R.A.} & \multicolumn{1}{c}{Dec.} & \multicolumn{1}{c}{Radio} & \multicolumn{1}{c}{X-ray} & \multicolumn{1}{c}{$L_{\rm 1.4~GHz}/$} & \multicolumn{1}{c}{$L_{\rm X}/$} & \multicolumn{1}{c}{$M_{200c}/$} & \multicolumn{1}{c}{$R_{200c}$} & \multicolumn{1}{c}{Rank} & \multicolumn{1}{c}{Class} \\
 & & \multicolumn{4}{c}{(deg., J2000.0)} & & & \multicolumn{1}{c}{$\rm W~Hz^{-1})$} & \multicolumn{1}{c}{$10^{42} \rm erg~s^{-1})$} & \multicolumn{1}{c}{$10^{13} \rm M_{\odot})$} & \multicolumn{1}{c}{(deg.)} \\
 \multicolumn{1}{c}{(1)} & \multicolumn{1}{c}{(2)}  & \multicolumn{1}{c}{(3)} & \multicolumn{1}{c}{(4)}& \multicolumn{1}{c}{(5)}& \multicolumn{1}{c}{(6)} & \multicolumn{1}{c}{(7)} & \multicolumn{1}{c}{(8)} & \multicolumn{1}{c}{(9)} & \multicolumn{1}{c}{(10)} & \multicolumn{1}{c}{(11)} & \multicolumn{1}{c}{(12)} & \multicolumn{1}{c}{(13)} & \multicolumn{1}{c}{(14)}\\
\hline\hline
2456  &  4  &  149.678007  &  1.550514  &  149.67828  &  1.530020  &  0.630  &  0.653$^{ p }$  &  23.13  &  2.31 $\pm$ 1.42  &  1.98 $\pm$ 0.89  &  0.0370  &  0  & AGN \\ 
2486  &  25  &  149.885327  &  1.753499  &  149.85402  &  1.770230  &  0.141  &  0.124$^{ s }$  &  21.49  &  1.58 $\pm$ 0.08  &  1.65 $\pm$ -0.05  &  0.0884  &  20  & SFG \\ 
2500  &  304  &  150.384362  &  2.391270  &  150.41328  &  2.409970  &  0.374  &  0.385$^{ s }$  &  22.61  &  1.89 $\pm$ 0.74  &  1.75 $\pm$ 0.41  &  0.0371  &  6  & SFG \\
\hline
\end{tabular}
\tablefoot{(Continued)}
\end{sidewaystable*}

\addtocounter{table}{-1}

\begin{sidewaystable*}
\caption{Radio properties of Group Galaxies in COSMOS}     
\label{tab:radioprops}      
\centering          
\begin{tabular}{l l l l c c c c c c c c c l}     
\hline\hline
\multicolumn{2}{c}{ID}  & \multicolumn{2}{c}{radio} & \multicolumn{2}{c}{X-ray} & \multicolumn{2}{c}{Redshift $z$} & \multicolumn{1}{c}{$\rm log_{10}($}& \multicolumn{1}{c}{$\rm log_{10}($}& \multicolumn{1}{c}{$log_{10}($}& \multicolumn{1}{c}{}& \multicolumn{1}{c}{BGG} & \multicolumn{1}{c}{Radio}\\ 
\multicolumn{1}{c}{3~GHz} & \multicolumn{1}{c}{X-ray} & \multicolumn{1}{c}{R.A.} & \multicolumn{1}{c}{Dec.} & \multicolumn{1}{c}{R.A.} & \multicolumn{1}{c}{Dec.} & \multicolumn{1}{c}{Radio} & \multicolumn{1}{c}{X-ray} & \multicolumn{1}{c}{$L_{\rm 1.4~GHz}/$} & \multicolumn{1}{c}{$L_{\rm X}/$} & \multicolumn{1}{c}{$M_{200c}/$} & \multicolumn{1}{c}{$R_{200c}$} & \multicolumn{1}{c}{Rank} & \multicolumn{1}{c}{Class} \\
 & & \multicolumn{4}{c}{(deg., J2000.0)} & & & \multicolumn{1}{c}{$\rm W~Hz^{-1})$} & \multicolumn{1}{c}{$10^{42} \rm erg~s^{-1})$} & \multicolumn{1}{c}{$10^{13} \rm M_{\odot})$} & \multicolumn{1}{c}{(deg.)} \\
 \multicolumn{1}{c}{(1)} & \multicolumn{1}{c}{(2)}  & \multicolumn{1}{c}{(3)} & \multicolumn{1}{c}{(4)}& \multicolumn{1}{c}{(5)}& \multicolumn{1}{c}{(6)} & \multicolumn{1}{c}{(7)} & \multicolumn{1}{c}{(8)} & \multicolumn{1}{c}{(9)} & \multicolumn{1}{c}{(10)} & \multicolumn{1}{c}{(11)} & \multicolumn{1}{c}{(12)} & \multicolumn{1}{c}{(13)} & \multicolumn{1}{c}{(14)}\\
\hline\hline
3556  &  186  &  150.211565  &  2.401837  &  150.21205  &  2.402810  &  0.904  &  0.904$^{ s }$  &  23.15  &  2.20 $\pm$ 1.44  &  1.75 $\pm$ 0.78  &  0.0204  &  2  & SFG \\ 
3581  &  217  &  149.997322  &  2.449265  &  150.0037  &  2.454870  &  0.759  &  0.732$^{ s }$  &  23.36  &  1.73 $\pm$ 1.15  &  1.52 $\pm$ 0.73  &  0.0196  &  7  & AGN \\ 
3643  &  128  &  150.575167  &  2.162272  &  150.58504  &  2.182050  &  0.533  &  0.543$^{ s }$  &  22.61  &  1.67 $\pm$ 1.02  &  1.55 $\pm$ 0.69  &  0.0242  &  2  & AGN \\
\hline
\end{tabular}
\tablefoot{(Continued)}
\end{sidewaystable*}

\addtocounter{table}{-1}

\begin{sidewaystable*}
\caption{Radio properties of Group Galaxies in COSMOS}     
\label{tab:radioprops}      
\centering          
\begin{tabular}{l l l l c c c c c c c c c l}     
\hline\hline
\multicolumn{2}{c}{ID}  & \multicolumn{2}{c}{radio} & \multicolumn{2}{c}{X-ray} & \multicolumn{2}{c}{Redshift $z$} & \multicolumn{1}{c}{$\rm log_{10}($}& \multicolumn{1}{c}{$\rm log_{10}($}& \multicolumn{1}{c}{$log_{10}($}& \multicolumn{1}{c}{}& \multicolumn{1}{c}{BGG} & \multicolumn{1}{c}{Radio}\\ 
\multicolumn{1}{c}{3~GHz} & \multicolumn{1}{c}{X-ray} & \multicolumn{1}{c}{R.A.} & \multicolumn{1}{c}{Dec.} & \multicolumn{1}{c}{R.A.} & \multicolumn{1}{c}{Dec.} & \multicolumn{1}{c}{Radio} & \multicolumn{1}{c}{X-ray} & \multicolumn{1}{c}{$L_{\rm 1.4~GHz}/$} & \multicolumn{1}{c}{$L_{\rm X}/$} & \multicolumn{1}{c}{$M_{200c}/$} & \multicolumn{1}{c}{$R_{200c}$} & \multicolumn{1}{c}{Rank} & \multicolumn{1}{c}{Class} \\
 & & \multicolumn{4}{c}{(deg., J2000.0)} & & & \multicolumn{1}{c}{$\rm W~Hz^{-1})$} & \multicolumn{1}{c}{$10^{42} \rm erg~s^{-1})$} & \multicolumn{1}{c}{$10^{13} \rm M_{\odot})$} & \multicolumn{1}{c}{(deg.)} \\
 \multicolumn{1}{c}{(1)} & \multicolumn{1}{c}{(2)}  & \multicolumn{1}{c}{(3)} & \multicolumn{1}{c}{(4)}& \multicolumn{1}{c}{(5)}& \multicolumn{1}{c}{(6)} & \multicolumn{1}{c}{(7)} & \multicolumn{1}{c}{(8)} & \multicolumn{1}{c}{(9)} & \multicolumn{1}{c}{(10)} & \multicolumn{1}{c}{(11)} & \multicolumn{1}{c}{(12)} & \multicolumn{1}{c}{(13)} & \multicolumn{1}{c}{(14)}\\
\hline\hline
4962  &  76  &  150.425649  &  2.021020  &  150.4221  &  2.007470  &  0.860  &  0.870$^{ s }$  &  23.36  &  2.13 $\pm$ 1.50  &  1.73 $\pm$ 0.89  &  0.0208  &  5  & SFG \\ 
4991  &  65  &  149.894388  &  1.947530  &  149.89253  &  1.944590  &  0.772  &  0.773$^{ s }$  &  23.29  &  1.66 $\pm$ 1.27  &  1.46 $\pm$ 0.84  &  0.0181  &  4  & SFG \\ 
5007  &  304  &  150.405059  &  2.410020  &  150.41328  &  2.409970  &  0.374  &  0.385$^{ s }$  &  22.38  &  1.89 $\pm$ 0.74  &  1.75 $\pm$ 0.41  &  0.0371  &  10  & SFG \\
\hline
\end{tabular}
\tablefoot{(Continued)}
\end{sidewaystable*}

\addtocounter{table}{-1}

\begin{sidewaystable*}
\caption{Radio properties of Group Galaxies in COSMOS}     
\label{tab:radioprops}      
\centering          
\begin{tabular}{l l l l c c c c c c c c c l}     
\hline\hline
\multicolumn{2}{c}{ID}  & \multicolumn{2}{c}{radio} & \multicolumn{2}{c}{X-ray} & \multicolumn{2}{c}{Redshift $z$} & \multicolumn{1}{c}{$\rm log_{10}($}& \multicolumn{1}{c}{$\rm log_{10}($}& \multicolumn{1}{c}{$log_{10}($}& \multicolumn{1}{c}{}& \multicolumn{1}{c}{BGG} & \multicolumn{1}{c}{Radio}\\ 
\multicolumn{1}{c}{3~GHz} & \multicolumn{1}{c}{X-ray} & \multicolumn{1}{c}{R.A.} & \multicolumn{1}{c}{Dec.} & \multicolumn{1}{c}{R.A.} & \multicolumn{1}{c}{Dec.} & \multicolumn{1}{c}{Radio} & \multicolumn{1}{c}{X-ray} & \multicolumn{1}{c}{$L_{\rm 1.4~GHz}/$} & \multicolumn{1}{c}{$L_{\rm X}/$} & \multicolumn{1}{c}{$M_{200c}/$} & \multicolumn{1}{c}{$R_{200c}$} & \multicolumn{1}{c}{Rank} & \multicolumn{1}{c}{Class} \\
 & & \multicolumn{4}{c}{(deg., J2000.0)} & & & \multicolumn{1}{c}{$\rm W~Hz^{-1})$} & \multicolumn{1}{c}{$10^{42} \rm erg~s^{-1})$} & \multicolumn{1}{c}{$10^{13} \rm M_{\odot})$} & \multicolumn{1}{c}{(deg.)} \\
 \multicolumn{1}{c}{(1)} & \multicolumn{1}{c}{(2)}  & \multicolumn{1}{c}{(3)} & \multicolumn{1}{c}{(4)}& \multicolumn{1}{c}{(5)}& \multicolumn{1}{c}{(6)} & \multicolumn{1}{c}{(7)} & \multicolumn{1}{c}{(8)} & \multicolumn{1}{c}{(9)} & \multicolumn{1}{c}{(10)} & \multicolumn{1}{c}{(11)} & \multicolumn{1}{c}{(12)} & \multicolumn{1}{c}{(13)} & \multicolumn{1}{c}{(14)}\\
\hline\hline
6567  &  11  &  150.203909  &  1.638906  &  150.19728  &  1.658950  &  0.217  &  0.220$^{ s }$  &  21.56  &  2.54 $\pm$ 0.75  &  2.23 $\pm$ 0.24  &  0.0838  &  2  & SFG \\ 
6570  &  395  &  149.826096  &  2.135200  &  149.8261  &  2.135200  &  0.168  &  0.169$^{ s }$  &  21.30  &  0.09 $\pm$ -0.10  &  0.67 $\pm$ 0.24  &  0.0314  &  0  & SFG \\ 
6674  &  199  &  150.708047  &  2.292326  &  150.70682  &  2.292530  &  1.099  &  1.095$^{ s }$  &  23.46  &  2.01 $\pm$ 1.66  &  1.70 $\pm$ 1.13  &  0.0226  &  41  & AGN \\
\hline
\end{tabular}
\tablefoot{(Continued)}
\end{sidewaystable*}

\addtocounter{table}{-1}

\begin{sidewaystable*}
\caption{Radio properties of Group Galaxies in COSMOS}     
\label{tab:radioprops}      
\centering          
\begin{tabular}{l l l l c c c c c c c c c l}     
\hline\hline
\multicolumn{2}{c}{ID}  & \multicolumn{2}{c}{radio} & \multicolumn{2}{c}{X-ray} & \multicolumn{2}{c}{Redshift $z$} & \multicolumn{1}{c}{$\rm log_{10}($}& \multicolumn{1}{c}{$\rm log_{10}($}& \multicolumn{1}{c}{$log_{10}($}& \multicolumn{1}{c}{}& \multicolumn{1}{c}{BGG} & \multicolumn{1}{c}{Radio}\\ 
\multicolumn{1}{c}{3~GHz} & \multicolumn{1}{c}{X-ray} & \multicolumn{1}{c}{R.A.} & \multicolumn{1}{c}{Dec.} & \multicolumn{1}{c}{R.A.} & \multicolumn{1}{c}{Dec.} & \multicolumn{1}{c}{Radio} & \multicolumn{1}{c}{X-ray} & \multicolumn{1}{c}{$L_{\rm 1.4~GHz}/$} & \multicolumn{1}{c}{$L_{\rm X}/$} & \multicolumn{1}{c}{$M_{200c}/$} & \multicolumn{1}{c}{$R_{200c}$} & \multicolumn{1}{c}{Rank} & \multicolumn{1}{c}{Class} \\
 & & \multicolumn{4}{c}{(deg., J2000.0)} & & & \multicolumn{1}{c}{$\rm W~Hz^{-1})$} & \multicolumn{1}{c}{$10^{42} \rm erg~s^{-1})$} & \multicolumn{1}{c}{$10^{13} \rm M_{\odot})$} & \multicolumn{1}{c}{(deg.)} \\
 \multicolumn{1}{c}{(1)} & \multicolumn{1}{c}{(2)}  & \multicolumn{1}{c}{(3)} & \multicolumn{1}{c}{(4)}& \multicolumn{1}{c}{(5)}& \multicolumn{1}{c}{(6)} & \multicolumn{1}{c}{(7)} & \multicolumn{1}{c}{(8)} & \multicolumn{1}{c}{(9)} & \multicolumn{1}{c}{(10)} & \multicolumn{1}{c}{(11)} & \multicolumn{1}{c}{(12)} & \multicolumn{1}{c}{(13)} & \multicolumn{1}{c}{(14)}\\
\hline\hline
7894  &  275  &  149.839301  &  2.693789  &  149.83842  &  2.675170  &  0.270  &  0.264$^{ s }$  &  21.72  &  1.28 $\pm$ 0.82  &  1.41 $\pm$ 0.73  &  0.0388  &  2  & SFG \\ 
7920  &  243  &  150.259569  &  2.769150  &  150.26115  &  2.768570  &  1.308  &  1.315$^{ p }$  &  23.27  &  2.52 $\pm$ 1.96  &  1.80 $\pm$ 1.03  &  0.0168  &  3  & SFG \\ 
7938  &  239  &  150.09445  &  2.734668  &  150.10533  &  2.723920  &  0.739  &  0.731$^{ s }$  &  22.74  &  1.87 $\pm$ 1.33  &  1.61 $\pm$ 0.85  &  0.0211  &  5  & SFG \\
\hline
\end{tabular}
\tablefoot{(Continued)}
\end{sidewaystable*}

\addtocounter{table}{-1}

\begin{sidewaystable*}
\caption{Radio properties of Group Galaxies in COSMOS}     
\label{tab:radioprops}      
\centering          
\begin{tabular}{l l l l c c c c c c c c c l}     
\hline\hline
\multicolumn{2}{c}{ID}  & \multicolumn{2}{c}{radio} & \multicolumn{2}{c}{X-ray} & \multicolumn{2}{c}{Redshift $z$} & \multicolumn{1}{c}{$\rm log_{10}($}& \multicolumn{1}{c}{$\rm log_{10}($}& \multicolumn{1}{c}{$log_{10}($}& \multicolumn{1}{c}{}& \multicolumn{1}{c}{BGG} & \multicolumn{1}{c}{Radio}\\ 
\multicolumn{1}{c}{3~GHz} & \multicolumn{1}{c}{X-ray} & \multicolumn{1}{c}{R.A.} & \multicolumn{1}{c}{Dec.} & \multicolumn{1}{c}{R.A.} & \multicolumn{1}{c}{Dec.} & \multicolumn{1}{c}{Radio} & \multicolumn{1}{c}{X-ray} & \multicolumn{1}{c}{$L_{\rm 1.4~GHz}/$} & \multicolumn{1}{c}{$L_{\rm X}/$} & \multicolumn{1}{c}{$M_{200c}/$} & \multicolumn{1}{c}{$R_{200c}$} & \multicolumn{1}{c}{Rank} & \multicolumn{1}{c}{Class} \\
 & & \multicolumn{4}{c}{(deg., J2000.0)} & & & \multicolumn{1}{c}{$\rm W~Hz^{-1})$} & \multicolumn{1}{c}{$10^{42} \rm erg~s^{-1})$} & \multicolumn{1}{c}{$10^{13} \rm M_{\odot})$} & \multicolumn{1}{c}{(deg.)} \\
 \multicolumn{1}{c}{(1)} & \multicolumn{1}{c}{(2)}  & \multicolumn{1}{c}{(3)} & \multicolumn{1}{c}{(4)}& \multicolumn{1}{c}{(5)}& \multicolumn{1}{c}{(6)} & \multicolumn{1}{c}{(7)} & \multicolumn{1}{c}{(8)} & \multicolumn{1}{c}{(9)} & \multicolumn{1}{c}{(10)} & \multicolumn{1}{c}{(11)} & \multicolumn{1}{c}{(12)} & \multicolumn{1}{c}{(13)} & \multicolumn{1}{c}{(14)}\\
\hline\hline
9856  &  167  &  150.09321  &  2.320811  &  150.09583  &  2.299060  &  0.381  &  0.361$^{ s }$  &  21.95  &  1.24 $\pm$ 0.56  &  1.35 $\pm$ 0.46  &  0.0286  &  15  & SFG \\ 
9953  &  11  &  150.206921  &  1.647137  &  150.19728  &  1.658950  &  0.218  &  0.220$^{ s }$  &  21.45  &  2.54 $\pm$ 0.75  &  2.23 $\pm$ 0.24  &  0.0838  &  8  & SFG \\ 
9995  &  132  &  149.52106  &  2.168769  &  149.5124  &  2.182790  &  1.328  &  1.360$^{ p }$  &  23.35  &  2.43 $\pm$ 1.93  &  1.80 $\pm$ 1.09  &  0.0182  &  1  & SFG \\
\hline
\end{tabular}
\tablefoot{(Continued)}
\end{sidewaystable*}

\begin{table*}[h!]
\begin{center}
\caption{Radio luminosity functions of group galaxies obtained with the \vmax\ method. A halo mass cut, $M_{\rm 200c} > 10^{13.5} M_{\odot}$, was applied. Note: the error on $\Phi$ is in dex.}
\label{tab:lumfun_vmax}
\renewcommand{\arraystretch}{1.5}
\ifonecol \scriptsize \fi
\begin{tabular}[t]{c c c c c c}
\hline\hline
$z$ &$z_{\rm med}$ & $\log\lum$ &$log\lum$(med)& $\log\Phi$& N \\
& &$[\text{W}\,\text{Hz}^{-1}]$ &$ [\text{W}\,\text{Hz}^{-1}]$& $[\text{Mpc}^{-3}\,\text{dex}^{-1}]$ &\\
\hline\hline
0.07 $-$ 0.4  & 0.345   & 21.268 $-$ 22.000  & 21.524  &$ -4.073 ^{ + 0.198 }_{ - 0.142 }$& 9  \\
&   & 22.000 $-$ 22.891  & 22.488  &$ -4.292 ^{ + 0.084 }_{ - 0.084 }$& 27  \\
&   & 22.891 $-$ 23.781  & 23.025  &$ -4.979 ^{ + 0.259 }_{ - 0.172 }$& 6  \\
&    & 23.781 $-$ 24.671  & 24.420  &$ -5.470 ^{ + 0.573 }_{ - 0.281} $& 2  \\
&   & 24.671 $-$ 25.587  & 25.459  &$ -5.181 ^{ + 0.343 }_{ - 0.208 }$& 4  \\
\hline
0.4 $-$ 0.7  & 0.566   & 22.082 $-$ 22.560  & 22.442  &$ -4.389 ^{ + 0.185 }_{ - 0.135 }$& 10  \\
&   & 22.560 $-$ 23.237  & 22.832  &$ -4.589 ^{ + 0.079 }_{ - 0.079 }$& 31  \\
&   & 23.237 $-$ 23.913  & 23.361  &$ -5.459 ^{ + 0.294 }_{ - 0.188 }$& 5  \\
&   & 23.913 $-$ 24.590  & 24.107  &$ -5.463 ^{ + 0.294 }_{ - 0.188 }$& 5  \\
&   & 24.590 $-$ 25.292  & 25.267  &$ -6.178 ^{ + 0.999 }_{ - 0.359 }$& 1  \\
\hline
0.7 $-$ 1.0  & 0.881   & 22.605 $-$ 22.920  & 22.791  &$ -4.118 ^{ + 0.109 }_{ - 0.109 }$& 22  \\
&  & 22.920 $-$ 23.715  & 23.298  &$ -4.678 ^{ + 0.061 }_{ - 0.061 }$& 51  \\
&  & 23.715 $-$ 24.510  & 24.056  &$ -5.254 ^{ + 0.109 }_{ - 0.109 }$& 16  \\
&  & 24.510 $-$ 25.304  & 24.566  &$ -5.983 ^{ + 0.422 }_{ - 0.236 }$& 3  \\
&   & 25.304 $-$ 26.125  & 26.099  &$ -6.475 ^{ + 0.999 }_{ - 0.359 }$& 1  \\
\hline
1.0 $-$ 1.6 &  1.201   & 23.083 $-$ 23.391  & 23.219  &$ -5.293 ^{ + 0.294 }_{ - 0.188 }$& 5  \\
& & 23.391 $-$ 24.487  & 23.601  &$ -5.794 ^{ + 0.113 }_{ - 0.113 }$& 15  \\
&  &  24.487 $-$ 25.584  & 25.019  &$ -6.753 ^{ + 0.573 }_{ - 0.281 }$& 2  \\
&  &  25.584 $-$ 26.707  & 26.680  &$ -7.064 ^{ + 0.999 }_{ - 0.359 }$& 1  \\
\hline
1.6 $-$ 2.3  & 1.987   & 23.746 $-$ 24.159  & 24.049  &$ -6.082 ^{ + 0.343 }_{ - 0.208 }$& 4  \\

\\
\hline
\end{tabular}
\end{center}
\end{table*}

\begin{table}[h!]
\begin{center}
\caption{Radio luminosity functions of BGGs obtained with the \vmax\ method. A halo mass cut, $M_{\rm 200c} > 10^{13.5} M_{\odot}$, was applied. Note: the error on $\Phi$ is in dex.}
\label{tab:lumfun_bgg}
\renewcommand{\arraystretch}{1.5}
\ifonecol \scriptsize \fi
\begin{tabular}[t]{c c c c c c}
\hline\hline
$z$ &$z_{\rm med}$ & $\log\lum$ &$log\lum$(med)& $\log\Phi$& N \\
& &$[\text{W}\,\text{Hz}^{-1}]$ &$ [\text{W}\,\text{Hz}^{-1}]$& $[\text{Mpc}^{-3}\,\text{dex}^{-1}]$ &\\
\hline\hline
0.07 $-$ 0.4 & 0.346   & 21.926 $-$ 22.000  & 21.948  &$ -4.506 ^{ + 0.999 }_{ - 0.359 }$& 1  \\
&   & 22.000 $-$ 22.889  & 22.522  &$ -5.051 ^{ + 0.294 }_{ - 0.188 }$& 5  \\
&   & 22.889 $-$ 23.779  & 23.159  &$ -5.270 ^{ + 0.422 }_{ - 0.236 }$& 3  \\
&   & 23.779 $-$ 24.668  & 24.361  &$ -5.771 ^{ + 0.999 }_{ - 0.359 }$& 1  \\
&   & 24.668 $-$ 25.583  & 25.557  &$ -5.783 ^{ + 0.999 }_{ - 0.359 }$& 1  \\
\hline
0.4 $-$ 0.7  & 0.617   & 22.560 $-$ 23.237  & 22.996  &$ -5.219 ^{ + 0.214 }_{ - 0.150 }$& 8  \\
&  & 23.237 $-$ 23.913  & 23.539  &$ -5.854 ^{ + 0.573 }_{ - 0.281 }$& 2  \\
&   & 23.913 $-$ 24.590  & 24.028  &$ -5.685 ^{ + 0.422 }_{ - 0.236 }$& 3  \\
&  & 24.590 $-$ 25.292  & 25.267  &$ -6.178 ^{ + 0.999 }_{ - 0.359 }$& 1  \\
\hline
0.7 $-$ 1.0  & 0.894   & 22.656 $-$ 22.920  & 22.761  &$ -4.911 ^{ + 0.422 }_{ - 0.236 }$& 3  \\
&   & 22.920 $-$ 23.715  & 23.338  &$ -5.316 ^{ + 0.121 }_{ - 0.121 }$& 13  \\
&   & 23.715 $-$ 24.510  & 24.095  &$ -5.506 ^{ + 0.198 }_{ - 0.142 }$& 9  \\
&   & 24.510 $-$ 25.304  & 24.566  &$ -6.461 ^{ + 0.999 }_{ - 0.359 }$& 1  \\
&   & 25.304 $-$ 26.125  & 26.099  &$ -6.475 ^{ + 0.999 }_{ - 0.359 }$& 1  \\
\hline
1.0 $-$ 1.6  & 1.161   & 23.083 $-$ 23.391  & 23.152  &$ -5.551 ^{ + 0.573 }_{ - 0.281 }$& 2  \\
&   & 23.391 $-$ 23.959  & 23.588  &$ -5.917 ^{ + 0.259 }_{ - 0.172 }$& 6  \\
&   & 23.959 $-$ 24.527  & 23.978  &$ -6.733 ^{ + 0.999 }_{ - 0.359 }$& 1  \\
&   & 24.527 $-$ 25.120  & 25.095  &$ -6.787 ^{ + 0.999 }_{ - 0.359 }$& 1  \\
\hline
1.6 $-$ 2.3  & 1.987   & 23.746 $-$ 24.159  & 24.049  &$ -6.082 ^{ + 0.343 }_{ - 0.208 }$& 4  \\
\\
\hline
\end{tabular}
\end{center}
\end{table}

\begin{table}
\begin{center}
\caption{Radio luminosity functions of satellites obtained with the \vmax\ method. A halo mass cut, $M_{\rm 200c} > 10^{13.5} M_{\odot}$, was applied. Note: the error on $\Phi$ is in dex.}
\label{tab:lumfun_sg}
\renewcommand{\arraystretch}{1.5}
\ifonecol \scriptsize \fi
\begin{tabular}[t]{c c c c c c}
\hline\hline
$z$ &$z_{\rm med}$ & $\log\lum$ &$log\lum$(med)& $\log\Phi$& N \\
& &$[\text{W}\,\text{Hz}^{-1}]$ &$ [\text{W}\,\text{Hz}^{-1}]$& $[\text{Mpc}^{-3}\,\text{dex}^{-1}]$ &\\
\hline\hline
0.07 $-$ 0.4 & 0.345   & 21.268 $-$ 22.000  & 21.522  &$ -4.090 ^{ + 0.214 }_{ - 0.150 }$& 8  \\
&   & 22.000 $-$ 22.891  & 22.432  &$ -4.375 ^{ + 0.093 }_{ - 0.093 }$& 22  \\
&  & 22.891 $-$ 23.781  & 22.933  &$ -5.289 ^{ + 0.422 }_{ - 0.236 }$& 3  \\
&   & 23.781 $-$ 24.671  & 24.478  &$ -5.771 ^{ + 0.999 }_{ - 0.359 }$& 1  \\
&   & 24.671 $-$ 25.587  & 25.360  &$ -5.306 ^{ + 0.422 }_{ - 0.236 }$& 3  \\
\hline
0.4 $-$ 0.7  & 0.530   & 22.082 $-$ 22.560  & 22.442  &$ -4.389 ^{ + 0.185 }_{ - 0.135 }$& 10  \\
&   & 22.560 $-$ 22.956  & 22.706  &$ -4.586 ^{ + 0.106 }_{ - 0.106 }$& 17  \\
&   & 22.956 $-$ 23.351  & 23.093  &$ -5.045 ^{ + 0.234 }_{ - 0.160 }$& 7  \\
&   & 23.351 $-$ 23.747  & 23.371  &$ -5.625 ^{ + 0.573 }_{ - 0.281 }$& 2  \\
&   & 23.747 $-$ 24.167  & 24.125  &$ -5.654 ^{ + 0.573 }_{ - 0.281 }$& 2  \\
\hline
0.7 $-$ 1.0  & 0.873   & 22.605 $-$ 22.920  & 22.796  &$ -4.181 ^{ + 0.119 }_{ - 0.119 }$& 19  \\
&   & 22.920 $-$ 23.480  & 23.191  &$ -4.683 ^{ + 0.075 }_{ - 0.075 }$& 34  \\
&   & 23.480 $-$ 24.040  & 23.720  &$ -5.378 ^{ + 0.214 }_{ - 0.150 }$& 8  \\
&   & 24.040 $-$ 24.600  & 24.391  &$ -5.707 ^{ + 0.343 }_{ - 0.208 }$& 4  \\
&   & 24.600 $-$ 25.185  & 25.160  &$ -6.327 ^{ + 0.999 }_{ - 0.359 }$& 1  \\
\hline
1.0 $-$ 1.6  & 1.230   & 23.196 $-$ 23.391  & 23.266  &$ -5.443 ^{ + 0.422 }_{ - 0.236 }$& 3  \\
&   & 23.391 $-$ 24.213  & 23.521  &$ -5.929 ^{ + 0.214 }_{ - 0.150 }$& 8  \\
&   & 24.213 $-$ 25.035  & 24.943  &$ -6.929 ^{ + 0.999 }_{ - 0.359 }$& 1  \\
&   & 25.858 $-$ 26.707  & 26.680  &$ -6.942 ^{ + 0.999 }_{ - 0.359 }$& 1  \\
\\
\hline
\end{tabular}
\end{center}
\end{table}

\begin{table}[h!]
\begin{center}
\caption{Best power law fit parameters. A halo mass cut, $M_{\rm 200c} > 10^{13.5} M_{\odot}$, was applied.}
\label{tab:lf_fit_par}
\renewcommand{\arraystretch}{1.5}
\ifonecol \scriptsize \fi
\begin{tabular}[t]{c c c c c}
\hline\hline
sample & $z$ & $\log\Phi$* & $\gamma$& $\log\Phi$* ($\gamma=-0.75$)\\
& & $[\text{Mpc}^{-3}\,\text{dex}^{-1}]$ &&$[\text{Mpc}^{-3}\,\text{dex}^{-1}]$\\
\hline\hline
GGs  & 0.07-0.40 & $-5.289 ^{ + 0.079 }_{ - 0.089 }$& $-0.518 ^{ + 0.054 }_{ - 0.053 }$& $-5.710^{+0.036}_{-0.040}$\\
GGs  & 0.40-0.70 & $-5.975 ^{ + 0.116 }_{ - 0.122 }$& $-1.090 ^{ + 0.103 }_{ - 0.104 }$& $-5.649^{+0.033}_{0.037}$\\
GGs  & 0.70-1.0 & $-5.323 ^{ + 0.038 }_{ - 0.042 }$& $-0.953 ^{ + 0.054 }_{ - 0.057 }$& $-5.231^{+0.023}_{-0.025}$\\
GGs  & 1.0-1.6 & $-6.089 ^{ + 0.074 }_{ - 0.130 }$& $-0.777 ^{ + 0.148 }_{ - 0.300 }$& $-6.063^{+0.044}_{-0.049}$\\
GGs  & 1.6-2.3 & $-6.044 ^{ + 0.113 }_{ - 0.142 }$& $-1.503 ^{ + 1.056 }_{ - 1.008 }$& $-6.083^{+0.109}_{-0.151}$\\
\hline
BGGs  & 0.07-0.40 & $-5.613 ^{ + 0.125 }_{ - 0.212 }$& $-0.378 ^{ + 0.108 }_{ - 0.158 }$& $-6.101^{+0.068}_{-0.083}$\\
\hline
\end{tabular}
\end{center}
\end{table}

\begin{table}[h!]
\begin{center}
\caption{$\Phi$* fits using the scaled method. A halo mass cut, $M_{\rm 200c} > 10^{13.5} M_{\odot}$, was applied.}
\label{tab:lf_fit_par_scaled}
\renewcommand{\arraystretch}{1.5}
\ifonecol \scriptsize \fi
\begin{tabular}[t]{c c c c c}
\hline\hline
sample & $z$ & $\log\Phi$* (scaled) \\
& & $[\text{Mpc}^{-3}\,\text{dex}^{-1}]$ \\
\hline\hline
GGs  & 0.07-0.40 & $-6.132 ^{ + 0.038 }_{ - 0.038 }$\\
GGs  & 0.40-0.70 & $-6.022 ^{ + 0.032 }_{ - 0.035 }$\\
GGs  & 0.70-1.0 & $-5.546 ^{ + 0.024 }_{ - 0.023 }$\\
GGs  & 1.0-1.6 & $-6.314 ^{ + 0.046 }_{ - 0.048 }$\\
GGs  & 1.6-2.3 & $-6.012 ^{ + 0.112 }_{ - 0.112 }$\\
\hline
BGGs  & 0.07-0.40 & $ -6.630 ^{ + 0.079 }_{ - 0.077 }$\\
BGGs  & 0.40-0.70 & $-6.400 ^{ + 0.066 }_{ - 0.067 }$\\
BGGs  & 0.70-1.0 & $ -6.199 ^{ + 0.053 }_{ - 0.052 }$\\
BGGs  & 1.0-1.6 & $ -6.572 ^{ + 0.076 }_{ - 0.073 }$\\
BGGs  & 1.6-2.3 & $ -6.004 ^{ + 0.106 }_{ - 0.108 }$\\
\hline
SGs  & 0.07-0.40 & $ -6.255 ^{ + 0.043 }_{ - 0.043 }$\\
SGs  & 0.40-0.70 & $ -6.096 ^{ + 0.039 }_{ - 0.039 }$\\
SGs  & 0.70-1.0 & $ -5.681 ^{ + 0.029 }_{ - 0.029 }$\\
SGs  & 1.0-1.6 & $ -6.555 ^{ + 0.068 }_{ - 0.065 }$\\
\hline
High-mass groups  & 0.07-0.40 & $ -6.132 ^{ + 0.038 }_{ - 0.038 }$\\
High-mass groups  & 0.40-0.70 & $ -6.021 ^{ + 0.034 }_{ - 0.033 }$\\
High-mass groups  & 0.70-1.0 & $ -5.546 ^{ + 0.024 }_{ - 0.024 }$\\
High-mass groups  & 1.0-1.6 & $ -6.317 ^{ + 0.049 }_{ - 0.049 }$\\
High-mass groups  & 1.6-2.3 & $ -6.010 ^{ + 0.116 }_{ - 0.109 }$\\
\hline
Low-mass groups  & 0.07-0.40 & $ -5.892 ^{ + 0.033 }_{ - 0.029 }$\\
Low-mass groups  & 0.40-0.70 & $ -6.746 ^{ + 0.073 }_{ - 0.077 }$\\
Low-mass groups  & 0.70-1.0 & $ -6.283 ^{ + 0.076 }_{ - 0.073 }$\\
\\
\hline
\end{tabular}
\end{center}
\end{table}

\begin{table}[h!]
\begin{center}
\caption{Radio luminosity functions of AGN inside X-ray galaxy groups obtained with the \vmax\ method. A halo mass cut, $M_{\rm 200c} > 10^{13.5} M_{\odot}$, was applied. Note: the error on $\Phi$ is in dex.}
\label{tab:lumfun_agn}
\renewcommand{\arraystretch}{1.5}
\ifonecol \scriptsize \fi
\begin{tabular}[t]{c c c c c c}
\hline\hline
$z$ &$z_{\rm med}$ & $\log\lum$ &$log\lum$(med)& $\log\Phi$& N \\
& &$[\text{W}\,\text{Hz}^{-1}]$ &$ [\text{W}\,\text{Hz}^{-1}]$& $[\text{Mpc}^{-3}\,\text{dex}^{-1}]$ &\\
\hline\hline

0.07 $-$0.4 & 0.344 &  22.820 $-$ 23.607 & 23.089 & -4.878 $^{+ 0.108}_{-0.108}$ & 16 \\
& &  23.607 $-$ 24.393 & 24.000 & -5.440 $^{+ 0.343}_{- 0.207 }$ & 4 \\
& & 24.393 $-$ 25.179 & 24.827 & -5.157 $^{+ 0.572}_{- 0.280 }$ & 2 \\
& & 25.179 $-$ 25.966 & 25.828 & -5.422 $^{+ 0.343}_{- 0.207}$ & 4 \\
\hline
0.4 $-$0.7 & 0.599 & 22.918 $-$ 23.587 & 23.241 & -5.992 $^{+ 0.233}_{- 0.160}$ & 7 \\
& & 23.587 $-$ 24.257 & 23.755 & -5.696 $^{+ 0.198}_{- 0.142}$ & 9 \\
& & 24.257 $-$ 24.926 & 24.491 & -5.817 $^{+ 0.259}_{- 0.172}$ & 6 \\
& & 24.926 $-$ 25.596 & 25.070 & -6.161 $^{+ 0.343}_{- 0.207}$ & 4 \\
\hline
0.7 $-$1.0 & 0.880 & 23.493 $-$ 24.204 & 23.867 & -6.069 $^{+ 0.566}_{- 0.365}$ & 19 \\
& & 24.204 $-$ 24.916 & 24.511 & -6.077 $^{+ 0.102}_{- 0.102}$ & 18 \\
& & 24.916 $-$ 25.628 & 25.342 & -6.368 $^{+ 0.233}_{- 0.160}$ & 7 \\
& & 25.628 $-$ 26.340 & 26.340 & -8.998 $^{+ 0.999 }_{- 0.359 }$ & 1 \\
\hline
1.0 $-$1.6 & 1.159 & 24.155 $-$ 24.961 & 24.565 & -7.352 $^{+ 0.259 }_{- 0.172 }$ & 6 \\
& & 24.961 $-$ 25.767 & 25.457 & -7.549 $^{+ 0.422 }_{- 0.236 }$ & 3 \\
& & 26.573 $-$ 27.379 & 27.379 & -9.823 $^{+ 0.999 }_{- 0.359 }$ & 1 \\
\\
\hline
\end{tabular}
\end{center}
\end{table}

\begin{table}[h!]
\begin{center}
\caption{Radio luminosity functions of SFGs inside X-ray galaxy groups obtained with the \vmax\ method. A halo mass cut, $M_{\rm 200c} > 10^{13.5} M_{\odot}$, was applied. Note: the error on $\Phi$ is in dex.}
\label{tab:lumfun_sfg}
\renewcommand{\arraystretch}{1.5}
\ifonecol \scriptsize \fi
\begin{tabular}[t]{c c c c c c}
\hline\hline
$z$ &$z_{\rm med}$ & $\log\lum$ &$log\lum$(med)& $\log\Phi$& N \\
& &$[\text{W}\,\text{Hz}^{-1}]$ &$ [\text{W}\,\text{Hz}^{-1}]$& $[\text{Mpc}^{-3}\,\text{dex}^{-1}]$ &\\
\hline\hline

0.07 $-$0.4 & 0.339 &  21.654 $-$ 22.354 & 21.992 & -4.784 $^{+ 0.125}_{- 0.125}$ & 12 \\
& &  22.354 $-$ 23.055 & 22.694 & -4.594 $^{+ 0.092}_{- 0.092}$ & 22 \\
& & 23.055 $-$ 23.756 & 23.232 & -4.740 $^{+ 0.198}_{- 0.142}$ & 9 \\
& & 23.756 $-$ 24.456 & 24.202 & -4.800 $^{+ 0.572}_{- 0.280}$ & 2 \\
\hline
0.4 $-$0.7 & 0.611 & 22.697 $-$ 23.262 & 23.147 & -5.647 $^{+ 0.120}_{- 0.120}$ & 13 \\
& & 23.262 $-$ 23.828 & 23.441 & -5.358 $^{+ 0.105}_{- 0.105}$ & 17 \\
& & 23.828 $-$ 24.393 & 24.044 & -5.312 $^{+ 0.572}_{- 0.280}$ & 2 \\
& & 24.393 $-$ 24.959 & 24.959 & -8.282 $^{+ 0.999}_{- 0.359}$ & 1 \\
\hline
0.7 $-$1.0 & 0.879 & 23.268 $-$ 23.709 & 23.457 & -5.607 $^{+ 0.083}_{- 0.083}$ & 27 \\
& & 23.709 $-$ 24.150 & 23.842 & -5.639 $^{+ 0.094}_{- 0.094}$ & 21 \\
& & 24.150 $-$ 24.592 & 24.244 & -5.391 $^{+ 0.233}_{- 0.160}$ & 7 \\
& & 24.592 $-$ 25.033 & 25.033 & -7.879 $^{+ 0.999}_{- 0.359}$ & 1 \\
\hline
1.0 $-$1.6 & 1.230 & 23.792 $-$ 24.116 & 24.023 & -6.954 $^{+ 0.669 }_{- 0.326 }$ & 8 \\
& & 24.116 $-$ 24.441 & 24.292 & -6.752 $^{+ 0.233}_{- 0.160}$ & 7 \\
& & 24.441 $-$ 24.765 & 24.695 & -7.016 $^{+ 0.422}_{- 0.236}$ & 3 \\
& & 24.765 $-$ 25.090 & 25.090 & -9.541 $^{+ 0.999}_{- 0.359}$ & 1 \\
\hline 
1.6 $-$ 2.3 & 1.987 & 24.602 $-$ 24.767 & 24.684 & -7.742 $^{+ 0.572}_{- 0.280 }$ & 2 \\
& & 24.767 $-$ 24.931 & 24.914 & -7.031 $^{+ 0.572 }_{- 0.280}$ & 2 \\\\
\hline
\end{tabular}
\end{center}
\end{table}

\begin{table}[h!]
\begin{center}
\caption{$\chi^{2}$ test results for the GGs (Sec.~\ref{sec:methodscomparison}), and AGN and SFGs (Sec.~\ref{sec:agn}). PL is for the power law, linear regression fit and SC for the scaled fit. DoF denotes the degrees of freedom.
}
\label{tab:chi2_agn_sfg}
\renewcommand{\arraystretch}{1.5}
\ifonecol \scriptsize \fi
\begin{tabular}[t]{c | c | c c c | c c c | c c c}
\hline\hline
& & \multicolumn{3}{c}{GGs} & \multicolumn{3}{c}{AGN} & \multicolumn{3}{c}{SFGs}\\
$z_{\rm bin}$ & model &$\chi^{2}$ & $p$-value & DoF &$\chi^{2}$ & $p$-value & DoF& $\chi^{2}$ & $p$-value & DoF \\
\hline\hline
1 & PL & 49.39  &  4e-10 & 4 & 11.54 & 0.009 & 3 & 4.37 & 0.224 & 3  \\
1 & SC & 72.74  &  5e-15 & 4 & 9.24 & 0.026 & 3 & 7.96 & 0.047 & 3 \\ 
2 & PL & 9.87  &  0.043 & 4 & 3.74 & 0.291 & 3 & 6.25 & 0.100 & 3  \\ 
2 & SC & 13.05  &  0.011 & 4 & 8.15 & 0.043 & 3 & 5.96 & 0.114 & 3  \\ 
3 & PL & 6.22  &  0.183 & 4 & 9.65 & 0.022 & 3 & 8.16 & 0.043 & 3  \\ 
3 & SC & 16.07  &  0.003 & 4 & 10.85 & 0.013 & 3 & 6.87 & 0.076 & 3  \\ 
4 & PL & 7.51  &  0.057 & 3 & 5.11 & 0.078 & 2 & 12.57 & 0.006 & 3  \\ 
4 & SC & 24.12  &  2e-5 & 3 & 5.33 & 0.070 & 2 & 9.15 & 0.027 & 3  \\ 
5 & PL & 0.00  &  $-$ & 0 & $-$ & $-$ & $-$ & 3.37 & 0.066 & 1  \\ 
5 & SC & 0.00  &  $-$ & 0 & $-$ & $-$ & $-$ & 1.86 & 0.172 & 1  \\  \\
\hline
\end{tabular}
\end{center}
\end{table}

\end{appendix}

\end{document}